\documentclass[review,authoryear,3p,times]{elsarticle}

\usepackage{lineno,hyperref}
\modulolinenumbers[1]

\newlength\savedwidth

\newcommand\thickhline{\noalign{\global\savedwidth\arrayrulewidth\global\arrayrulewidth 2pt}%
\hline
\noalign{\global\arrayrulewidth\savedwidth}}

\usepackage{xspace}
\usepackage{soul}
\usepackage{cancel}
\usepackage{mathtools}
\usepackage{makecell}
\usepackage{nicefrac}
\usepackage{float}
\usepackage{wrapfig}
\usepackage{xcolor}

\newcommand{\bs}[1]{\ensuremath{\boldsymbol{#1}}}

\definecolor{dgreen}{rgb}{0.0, 0.5, 0.0}

\usepackage{textcomp}

\RequirePackage[normalem]{ulem} 
\RequirePackage{color}\definecolor{RED}{rgb}{1,0,0}\definecolor{BLUE}{rgb}{0,0,1} 
\usepackage{siunitx}
\usepackage{longtable}
\usepackage{multirow}
\usepackage[inline]{enumitem}
\usepackage{cleveref}
\crefname{equation}{}{}
\allowdisplaybreaks

\journal{Computers and Electronics in Agriculture}

\let\oldequation\equation
\let\oldendequation\endequation

\renewenvironment{equation}
  {\linenomathNonumbers\oldequation}
  {\oldendequation\endlinenomath}

\let\oldalign\align
\let\oldendalign\endalign

\renewenvironment{align}
  {\linenomathNonumbers\oldalign}
  {\oldendalign\endlinenomath}
\let\oldalignat\alignat
\let\oldendalignat\endalignat

\renewenvironment{alignat}
  {\linenomathNonumbers\oldalignat}
  {\oldendalignat\endlinenomath}

\begin{document}

\begin{frontmatter}

\title{Modeling and optimal control of growth, energy, and resource dynamics of \textit{Hermetia illucens} in mass production environment}

\author{Murali Padmanabha}\ead{murali.padmanabha@etit.tu-chemnitz.de}
\author{Alexander Kobelski}
\author{Arne-Jens Hempel}
\author{Stefan Streif}\ead{stefan.streif@etit.tu-chemnitz.de}
\address{Automatic Control and System Dynamics Lab, Technische Universität Chemnitz, Chemnitz, 09107, Germany}

\begin{abstract}
Mass production of \textit{Hermetia illucens} insect larvae is now being adopted in many countries and is taking an industrial production approach.
Despite abundant literature on factors that affect larvae growth and the optimal static parameters identified in laboratory setup, for an industrial production process it is necessary to identify the trajectories such that the growth as well as the production process is optimal.
To achieve this in this work, some of the important requirements and challenges involved thereof are identified and objectives of the automation process are formulated within a model based optimal control setup.
Mechanistic models necessary for the optimization framework are derived as differential equations that describe the dynamic variation of resources (feed, water, O$_2$ etc.), energy, and larval biomass. 
In addition, the elevated metabolic activity of larvae corresponding to the final instar development is identified and also modelled based on the observation from experiments.
The mass and energy balance approach used in modelling enables the quantification and distinction of the mass and energy flux components in various levels (e.g. larvae body, growing medium, production environment, and external environment) while holding its applicability for both open and closed/reactor based production setups. 
Finally, the trajectories generated using the synthesized optimal controller are tested under different scenarios showcasing significant reduction in resource consumption compared to a fixed set-point operation of the production setup.
Results presented in this work not only showcase the potential of the mechanistic models and their application in identifying the relevant process parameters (e.g. reactor properties such as volume, thermal conductivity, actuator capacities), but most importantly in optimizing the process dynamically and tuning the process objectives as desired (e.g. maximize larvae mass, reduce energy). 


\end{abstract}


\end{frontmatter}

\graphicspath{{./gfx/pdf}}
\section{Introduction}
\textit{Hermetia illucens} also commonly known as Black soldier fly (BSF) is widely researched for its use in animal feed production and bio waste management and recycling processes. 
Their rapid growth and robust bio-waste conversion capabilities have gained attention and have proven \textit{Hermetia illucens} to be a suitable candidate for efficient biomass production.

Several studies have been performed to understand the requirements for both rearing of larvae for proteins and rearing of flies for reproduction \citep{Bondari1981,Newton1977,Tomberlin2009,Barragan-Fonseca2017}.
Some of these works have even proposed static models describing the growth rates as a function of growing conditions \citep{Chia2018,Gligorescu2019,Shumo2019,Palma2018,Padmanabha2020}.
A significant number of publications focussing on the waste conversion efficiencies of the BSF larvae can be found \citep{Diener2009,ParraPaz2015,Manurung2016,Gao2019,Lalander2019}.
More recent literatures show the transition of biomass conversion efficiency studies performed in labs to large scale production environments \cite{Miranda2020,Scala2020}.
These trends indicate that the mass production of insect larvae is now being adopted in many countries and is taking an industrial production approach.
Irrespective of the application of BSF larvae to produce high quality protein rich animal feed or to simply accelerate the reduction of biological waste, there are no studies or literature on the automation and control of the processes involved.

Some of the state of the art and relevant work which provide insights to the challenges involved in the production and facilitate the development of models are presented here.
The threshold temperatures and thermal requirements were highlighted by \cite{Chia2018}.
A comparison of the development rates over different temperature ranges was presented by \cite{Shumo2019}.
The effect of humidity on the egg eclosion and adult emergence were studied by \cite{Holmes2012}.
The influence of diet, its moisture content and the temperatures were presented together to showcase its importance in the development of the larvae \citep{Diener2009,Harnden2016,Guo-Hui2014,Gligorescu2018}.
The effects of moisture content in food waste on residue separation was presented by \cite{Cheng2017} showing the importance of the moisture in post-processing.
Another study by \cite{Meneguz2018}, showed the effects of pH levels of the substrate (feed), in which larvae are grown, on the larval development.
Despite a large number of literature on BSF production, studies on this process optimization is limited to identifying the optimal static condition such as optimal but fixed set-points for temperature \citep{Chia2018}, feeding rate \citep{Diener2009}, feed components \citep{Lee2021}, etc., that result in better biomass production.
In order to perform a dynamic optimization of the rearing process including the energy and resources involved, it is necessary to identify various aspects of the production process and derive dynamic models that describe these aspects.

Based on the state of the art literature and industrial practices, some of the challenges inherent to the automation and optimization of commercial production of the \textit{Hermetia illucens} larvae have been identified in this work. 
Firstly, high substrate temperature and moisture requirements pose the problem of increased evaporation. 
At lower temperatures, metabolic activity and thus the growth rate is reduced. 
Depending on food quality and larval density, the peak metabolic activity of the larvae generate heat and result in substrate temperatures up to \SI{45}{\celsius}. 
This requires cooling down the substrate to reduce the mortality rate of the larvae.
Secondly, the fast metabolic rates in larvae results in increased oxygen consumption and depending on the feed, the rates of carbon dioxide and ammonia production also increases. 
This requires frequent supply of fresh air which also accelerates the evaporation of water from the substrate.
Thirdly, the moisture content of the substrate has to be high enough to render it edible for the young larvae. 
Depending on the harvesting procedure, the substrate has to be dry to reliably separate the larvae from the substrate, which conflicts with the moisture requirements for the larvae growth.
Finally, for production of larvae under constrained resources, it is necessary to find the optimal growing conditions that can result in resource and/or cost savings.

In order to address the above identified challenges, the process involved and the corresponding model requirements are formulated as in Table~\ref{tab:reqs}.

\begin{table}[H]\caption{\bf Process v.s. model requirements} \label{tab:reqs}
\begin{tabular}{c m{0.45\textwidth}|m{0.45\textwidth}}
	\hline
	\textbf{\#} & Process & Model requirement \\ \thickhline
	\textbf{1.} &Dynamic temperature control regime to promote growth and development & Dynamic variation of temperature in substrate due to metabolic activity and heat exchange with the environment\\\hline
	\textbf{2.} &Ventilation strategies for growth promoting gas concentration and low toxic gas accumulation &  Dynamic gas concentration changes due to metabolic activity and exchange with environment\\ \hline
	\textbf{3.} &Water content regulation for moist substrate during growth phase and dry substrate for extraction & Dynamic change in substrate moisture content and evaporation\\\hline
	\textbf{4.} &Feeding strategies for optimal feed consumption and less waste generation & Dynamic change in nutrition content and accumulation of waste (faeces and exhausted feed) in substrate \\ \hline	
	\textbf{5.} &In-situ monitoring/estimation of larval growth and substrate properties & Dynamic change in individual substrate component (feed, larvae, water) masses \\\hline
	\textbf{6.} &Maximize larvae dry-mass production & Dynamic growth and development changes and its dependency on the influence of growing conditions\\\hline
\end{tabular}
\end{table}

The dynamic models presented in our previous work \citep{Padmanabha2020} fulfill only the requirement \textbf{6}, which is critical since the optimization depends on the larvae biomass changes.
To the best knowledge of the authors, there exists no other work that satisfies the other models requirements (\textbf{1}--\textbf{5}).
This also follows that no literature has addressed the automation strategies or the application of optimal control in the context of \textit{Hermetia illucens} production.
This work addresses both these aspects by firstly, defining the \textbf{optimal control} framework to meet the process objectives defined above and secondly, deriving all the necessary detailed \textbf{mechanistic models}.
The models derived in this work use the biomass growth and larvae development model presented in our previous work \citep{Padmanabha2020} to obtain the production and consumption of various energy and resource fluxes that depend on the evolving dry mass of the larvae.
Also, the data necessary to analyze the process and complete the models are obtained using our previously developed laboratory scale production environment \citep{Padmanabha2019}.
Along with the application in optimal control, the models presented in this work can also be used to perform simulation studies of the growth and production process, design of control policies for the large scale production, etc.

The following sections provide a detailed approach taken to define the control goals, develop the necessary models, analyze the experiment data sets, obtain the model parameters, and evaluate the performance of the implemented control framework.
Firstly, in Section~\ref{sec:production_system}, a typical production setup is described along with a mathematical notation of its system states and the flow of various resources. Also, an optimal control problem is formulated for the production process based on the desired control objectives.
This is followed by, in Section~\ref{sec:modelling}, the details on extending the models of the previous work and obtaining the resource and energy flux models for the production setup.
In Section~\ref{sec:data_tools}, the process used to obtain the experiment data and the mathematical tools used to solve the optimal control problem and the parameter estimation problems are described. 
The results of the model simulation are compared with the actual measurement data and the quality of fit is determined in Section~\ref{sec:results_model}.
Finally, in Section~\ref{sec:results_process_analysis} and \ref{sec:results_oc}, the application of the model derived in this work for process analysis and specifically for the optimization of resources using a model based controller is showcased.

\section{Materials and Methods}
In this section, an overview to a typical production setup is presented with all the state variables, control variables and disturbances or external resources.
Based on this, the setup of the optimal control framework is described followed by the detailed modelling of the resource and energy fluxes between various components in the production setup.
The experiments performed to study the process are described along with the setup used for the model parameter identification.

\subsection{Overview of the production setup and the requirements for automation and control}\label{sec:production_system}
To provide an overview of a production setup and to transition to the model development process, an overview of all the state variables, inputs, and disturbances or external resources are listed in Table~\ref{tab:states}.

\begin{longtable}{m{0.1\textwidth} m{0.7\textwidth} m{0.1\textwidth}}\caption{\textbf{List of symbols used to describe the states, disturbances, and inputs of the system.}\label{tab:states}}\\
		\hline	\bf	Symbol                     & \bf Description                                		& \bf Unit                  \\ \hline \endfirsthead
		\caption[]{(continued ...)} \\
		\hline	\bf	Symbol                     & \bf Description                                		& \bf Unit                  \\ \hline 	\endhead
		\multicolumn{3}{r}{{continued on next page ...}} \\ \hline	\endfoot
		\hline \hline \endlastfoot
		\multicolumn{3}{l}{State variables:}\\
		$L_\mathrm{num}$           & number of larvae in the growing medium & [-]  \\
        $T_\mathrm{\Sigma}$        & development sums used to track the developmental stage of larva       			& [\si{\hour}] \\
		$B_\mathrm{dry}$           & dry mass per larva 				        & [\si{\gram}] \\
		$B_\mathrm{wet}$           & wet mass per larva 				        & [\si{\gram}] \\
        $B_\mathrm{med}$           & combined mass of larvae, water and feed 	& [\si{\kilogram}] \\
        $B_\mathrm{tot}$           & total dry mass of all larvae in substrate  & [\si{\gram}] \\
        $N_\mathrm{med}$           & total substrate (feed and excreta) dry mass in medium 		& [\si{\kilogram}] \\
		$N_\mathrm{feed}$          & total unused feed in substrate		 		& [\si{\kilogram}] \\
		$N_\mathrm{exc}$           & larvae excreta present in substrate        & [\si{\kilogram}] \\
		$W_\mathrm{med}$           & total water in the growing medium          & [\si{\kilogram}] \\
		$T_\mathrm{med}$           & temperature of growing medium   			& [\si{\celsius}] \\
		$T_\mathrm{air}$           & temperature of air in production environment 	& [\si{\celsius}] \\
		$C_\mathrm{air}$           & CO$_2$ concentration of air in production unit & [\si{\kilogram\per\meter\cubed}]\\
		$H_\mathrm{air}$           & absolute humidity of the air in production unit & [\si{\kilogram\per\meter\cubed}] \\
		$O_\mathrm{air}$           & O$_2$ concentration of air in production unit 	& [\si{\kilogram\per\meter\cubed}]  \\
		$T_\mathrm{chm}$           & temperature on surface of walls of production system/reactor & [\si{\celsius}] \\        
		$T_\mathrm{hx}~^+$            & temperature of the heat exchanger inside production system   & [\si{\celsius}] \\
        $W_\mathrm{chm}~^+$           & total water or condensate on the inner walls of the production environment & [\si{\kilogram}] \\
		$W_\mathrm{hx}~^+$           & total water or condensate on the surface of the heat exchanger & [\si{\kilogram}] \\
		\\
		\multicolumn{3}{l}{Disturbances or external resources:}\\
		$C_\mathrm{out}$           & CO$_2$ concentration of external source & [\si{\kilogram\per\meter\cubed}] \\
		$T_\mathrm{out}$           & temperature of outside air             & [\si{\celsius}] \\
		$H_\mathrm{out}$           & absolute humidity of external source   & [\si{\kilogram\per\meter\cubed}]  \\
		$W_\mathrm{out}$           & water in the external source         	& [\si{\kilogram}] \\
		$N_\mathrm{out}$           & Nutrients or feed in the external source         	& [\si{\kilogram}] \\
		$E_\mathrm{out}$           & (electrical) energy from the external source     	& [\si{\watt}] \\
		\\
		\multicolumn{3}{l}{Input variables:}\\
		$u_\mathrm{v}$ 				& input signal to ventilator pump & [-] \\
		$u_\mathrm{d}$ 				& input signal to opening of the door & [-] \\
		$u_\mathrm{H}$ 				& input signal to humidifier & [-] \\
		$u_\mathrm{T}$ 				& input signal to heater/cooler & [-] \\
		$u_\mathrm{N}$ 				& input signal to feeder & [-] \\
		$u_\mathrm{W_{med}}$ 		& input signal to water pump for growing medium & [-] \\
		$u_\mathrm{fan}$ 			& input signal to blower/fan for air flow & [-] \\
		$u_\mathrm{h}$ 				& removal/harvesting of larvae from substrate & [-] \\
		$u_\mathrm{W_{sto}}~^+$ 	& input signal to water pump for storage tank for humidifier & [-] \\
		$u_\mathrm{W_{ovf}}~^+$ 	& input signal to water pump for condensate removal & [-] \\
		$u_{\mathrm{I}i}~^+$		& input to LED channel number $i$ & [-]	\\
		\multicolumn{3}{l}{\footnotesize{$~^+$ Reactor specific variables}}
\end{longtable}\par

\begin{figure}[H]
	\includegraphics[width=\linewidth,trim={20 0 0 0}]{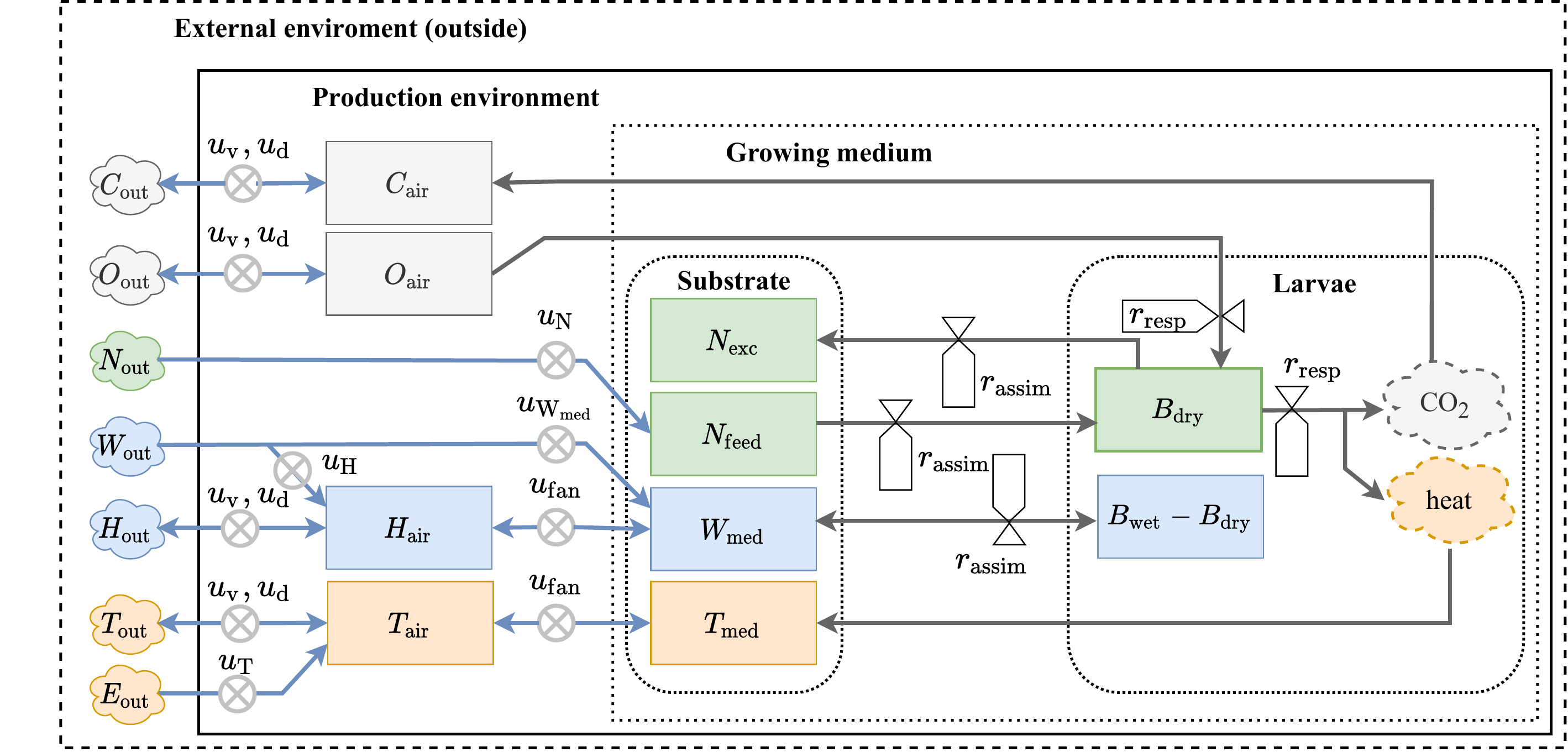}\caption{\textbf{Overview of the larvae production setup and the internal resource flows.} 
It consists of a growing medium filled with larvae and feed, placed in either a closed or an open production environment. 
The valves representing the rates labelled as $r_\mathrm{resp}$ and $r_\mathrm{assim}$ are internally regulated by the larvae and the substrate states. Circular valves labelled $u_\mathrm{xyz}$ represent the control signals that control the flow of resources (fluxes).}\label{fig:prod_ov}
\end{figure}

A typical production setup, as illustrated in Fig~\ref{fig:prod_ov}, consists of several growing trays filled with substrate made of feed and water and populated with young larvae neonates.
The growing trays consisting of the substrate and the larvae are referred to as growing medium.
The environment where the growing mediums are placed are referred to as production environment in this work.
The environment where the production setup is located and the resources are sourced is referred to as external or outside environment.

The production environment may be completely closed, partially closed, or fully open. In a completely closed setup, as in case of complex closed bio reactors, all the resources (air, heat, humidity, water) to the production environment are supplied from external sources or generated using actuators.
Such closed production setup can be described as
\begin{equation}\label{eq:prod_sys_1}
\begin{aligned}
	\bs{x} &= [B_\mathrm{dry}~B_\mathrm{wet}~T_\mathrm{\Sigma}~W_\mathrm{med}~N_\mathrm{feed}~N_\mathrm{exc}~B_\mathrm{med}~T_\mathrm{med}~T_\mathrm{air}~H_\mathrm{air}~C_\mathrm{air}~O_\mathrm{air}~T_\mathrm{chm}~T_\mathrm{hx}~W_\mathrm{chm}~W_\mathrm{hx}]^\top,\\
	\bs{u} &= [u_\mathrm{v}~u_\mathrm{d}~u_\mathrm{T}~u_\mathrm{H}~u_\mathrm{W_{med}}~u_\mathrm{N}~u_\mathrm{fan}~u_\mathrm{h}]^\top, \\
	\bs{d} &= [T_\mathrm{out}~H_\mathrm{out}~C_\mathrm{out}~O_\mathrm{out}]^\top,
\end{aligned}
\end{equation}
where $\bs{x}$ is a set of all state variables, $\bs{u}$ is a set of all input variables and $\bs{d}$ is a set of all disturbances or external resources.
In case of a partially closed setup as in case of large halls, resources available from external environment are supplied through ventilation and partly modified (heating, cooling, humidification, dehumidification) using actuators.
Such production setup are easy to establish and are less complex due to the fewer number of actuator compared to the closed setup and thus can only be partly automated/controlled.
Such production setups can be described as
\begin{equation}\label{eq:prod_sys_2}
	\begin{aligned}
		\bs{x} &= [B_\mathrm{dry}~B_\mathrm{wet}~T_\mathrm{\Sigma}~W_\mathrm{med}~N_\mathrm{feed}~N_\mathrm{exc}~B_\mathrm{med}~T_\mathrm{med}]^\top,\\
		\bs{u} &= [u_\mathrm{\Delta T}~u_\mathrm{\Delta H}~u_\mathrm{W_{med}}~u_\mathrm{N}]^\top,\\
		\bs{d} &= [T_\mathrm{out}~H_\mathrm{out}~C_\mathrm{out}~O_\mathrm{out}]^\top,
	\end{aligned}
	\end{equation}
with $u_\mathrm{\Delta T}$ and $u_\mathrm{\Delta H}$ representing the modification (addition or removal) done to the temperature $T_\mathrm{out}$ and humidity $H_\mathrm{out}$ of the external air such that $T_\mathrm{air} = T_\mathrm{out} + u_\mathrm{\Delta T}$ and $H_\mathrm{air} = H_\mathrm{out} + u_\mathrm{\Delta H}$.	
Finally, a completely open setup is where the external and production environment are the same and no additional actuator exist for the modification of the environment ($\bs{u} = [u_\mathrm{W_{med}}]$).
Based on the description of the different possible production setups, the optimal control task for automation and resource optimization can be formulated specifically.

\subsection{Model based controller synthesis for production process optimization}\label{sec:MPC_application}
To apply the optimal control framework to a process, the process goals need to be formulated as an objective function.
In this work we formulate the objective function and thus the optimal control problem based on the process goals listed in Table~\ref{tab:reqs}.
Such optimal control formulation are applied in greenhouses for automation and resource optimization \citep{vanStraten2011,Padmanabha2020a}. It is also widely accepted for control process optimization \citep{Zhang2020} and also applied in aquaculture and aquaponics for optimal fish growth and resource utilization \citep{Cacho1991,Karimanzira2017,Chahid2021}.

Given the models of the larvae dynamics, substrate dynamics, and the surrounding environment, the optimization/optimal control problem can be defined as
\begin{alignat}{2}
	\displaystyle \underset{\begin{array}{c}\bs{x}(\cdot), \bs{u}(\cdot)\end{array}}{\text{minimize}} \quad &-\alpha_1B_\mathrm{dry}(t_\mathrm{h}) + \alpha_2W_\mathrm{med}(t_\mathrm{h}) + \alpha_3N_\mathrm{med}(t_\mathrm{h}) +\int_{t_0}^{t_\mathrm{h}} \bs{u}(t)^\top ~\mathbf{R}~\bs{u}(t)  + \bs{\dot{u}}(t)^\top ~\mathbf{S} ~\bs{\dot{u}}(t)~dt &&\label{eq:oc_problem}\\
	\text{s.t.}\quad &\bs{\dot{x}}(t) = \bs{f}(\bs{x}(\mathit{t}),\bs{u}(\mathit{t}),\bs{d}(\mathit{t}),\mathbf{p}) \quad  t \in [t_0,t_\mathrm{h}], &&\text{system dynamics} \nonumber\\
				&\bs{{x}}(t_0) = \mathbf{x_0}, &&\text{initial conditions} \nonumber\\
				&\bs{x}(t) \in \mathcal{X} \quad \forall t \in [t_0,t_\mathrm{h}], &&\text{state constraints} \nonumber\\
				&\bs{u}(t) \in \mathcal{U} \quad\forall t \in [t_0,t_\mathrm{h}], &&\text{input constraints} \nonumber\\ 
				&\bs{\dot{u}}(t) \in \mathcal{V} \quad\forall t \in [t_0,t_\mathrm{h}], &&\text{input rate change constraints} \nonumber	
\end{alignat}
where $\alpha_1$, $\alpha_2$, and $\alpha_3$ correspond to the weights for adjusting the process goals of maximum dry-mass v.s. dry substrate v.s. minimum feed waste at harvest $t_\mathrm{h}$ respectively, $\mathbf{R}$ is a weight matrix that defines the cost of operation of the actuators for the entire production time, $\bs{\dot{u}}$ is the derivative of $\bs{u}$ with respect to time representing the change rate of the actuators signals, the matrix $\mathbf{S}$ defines the penalization of fast or sudden changes in the values of the actuators during the entire production time, $\bs{\dot{x}}$ the derivative of $\bs{x}$ with respect to time, $\mathbf{p}$ is the model parameter vector, $\mathcal{X}$, $\mathcal{U}$, and $\mathcal{V}$ are the sets that define the permitted values and bounds for the state variables, inputs, and input rate change respectively.
The values of $\alpha_1$, $\alpha_2$, $\alpha_3$ and $\mathbf{R}$ selected in this work do not correspond to actual monetary value but rather normalized values for tuning the resource utilization profile.
To solve the optimal control problem formulated by equation~\eqref{eq:oc_problem}, the system dynamics are necessary and will be derived in the following section.



\subsection{Modeling the larvae production process}\label{sec:modelling}
The models describing the growth and development of a larva presented by \cite{Padmanabha2020} are first used to obtain the models of the resource and energy flux generated by the larvae.
Using these flux models, the detailed models representing the different energy and resource fluxes as well as the dynamic changes of the states in a typical production setup are modelled.
Some of these flux terms that originate in the production environment are independent of the organism growing inside and thus allows for the reuse of models or parts of model equations for production of other organisms as well as other production setups.
In this work, the model equations presented in our previous work on lettuce production \citep{Padmanabha2020} are reused to derive the actuator contributed fluxes (e.g. TEC heating-cooling system and LEDs).
A list of all the energy and resources fluxes used for modelling are summarized in the Table~\ref{tab:fluxes} in \ref{app:Model_flux} for reference.

\subsubsection{Extension of the dynamic growth and development models of larvae}
The equation representing the dry mass change in the larvae as presented in \cite{Padmanabha2020} is given as
\begin{equation}\label{eq:dB_dry_plos/dt}
	\frac{\mathrm{d} B_\mathrm{dry}}{\mathrm{d}t}= \rlap{$\underbrace{\phantom{\phi_{\mathrm{B_{ing}}} - \phi_{\mathrm{B_{excr}}}  - \phi_{\mathrm{B_{assim}}}}}_{\phi_{\mathrm{B_{eff}}}}$} \phi_{\mathrm{B_{ing}}} - \phi_{\mathrm{B_{excr}}}  -
	\overbrace{\phi_{\mathrm{B_{assim}}} - \phi_{\mathrm{B_{mat}}} - \phi_{\mathrm{B_{maint}}}}^{\phi_{\mathrm{B_{metab}}}},
	\end{equation}
	where $\phi_{\mathrm{B_{ing}}}$ is feed flux to larvae, $\phi_{\mathrm{B_{excr}}}$ is the non digested feed flux back to the substrate, $\phi_{\mathrm{B_{assim}}}$ is feed spent in assimilation process, $\phi_{\mathrm{B_{maint}}}$ is the assimilates spent for basal maintenance of existing structure, $\phi_{\mathrm{B_{mat}}}$ is the assimilates spent for generating new structure and $\phi_{\mathrm{B_{metab}}}$ represents all the assimilates consumed for metabolic activities.

This equation~(4) of \cite{Padmanabha2020} combines both the maintenance and maturity as a single process due to data unavailability for validation.
However, in this work we explicitly model maturity and maintenance process and quantify the biomass flux $\phi_{\mathrm{B_{mat}}}$ responsible for the larvae final maturity as
\begin{align}\label{eq:dB_dry/dt_new}
	\frac{\mathrm{d}B_\mathrm{dry}}{\mathrm{d}t}&= \overbrace{\epsilon_\mathrm{inges} \ r_\mathrm{assim} \ k_\mathrm{inges} \ B_\mathrm{dry}}^{\phi_\mathrm{B_{eff}}} - \overbrace{\ r_\mathrm{maint} \ k_\mathrm{maint}\ B_\mathrm{dry}}^{\phi_\mathrm{B_{maint}}}- \overbrace{\ r_\mathrm{mat} \ k_\mathrm{mat}\ B_\mathrm{dry}}^{\phi_\mathrm{B_{mat}}},
\end{align}
where  $\epsilon_\mathrm{inges} = (1-k_\mathrm{\alpha_{excr}} - k_\mathrm{\alpha_{assim}}),$ represents the efficiency of the feed, and $r_\mathrm{assim}$, $r_\mathrm{maint}$, and $r_\mathrm{mat}$ being the assimilation rate, maintenance rate, and maturity rate functions respectively.
The rate function $r_\mathrm{assim}$ remains unchanged from the original work.
However, the maintenance and maturity rate functions are defined in this work as two separate rate functions.

The maintenance rate function is given as
\begin{align}\label{eq:r_lrv_maint}
	r_\mathrm{maint} &= \left( \dfrac{r_\mathrm{T}(T_\mathrm{med})}{k_\mathrm{r_{max}T}} \ \dfrac{r_\mathrm{F_{grw}}(B_\mathrm{feed})}{k_\mathrm{r_{max}gm}}  \ \dfrac{r_\mathrm{A}(A_\mathrm{air})}{k_\mathrm{r_{max}A}} \right),	\end{align}
	where $\dfrac{r_\mathrm{T}(T_\mathrm{med})}{k_\mathrm{r_{max}T}}$, $\dfrac{r_\mathrm{F_{grw}}(B_\mathrm{feed})}{k_\mathrm{r_{max}gm}}$, and $\dfrac{r_\mathrm{A}(A_\mathrm{air})}{k_\mathrm{r_{max}A}}$ are the normalized growth regulation rate functions dependent on the substrate temperature, feed, and air concentration respectively and $k_\mathrm{r_{max}T}$, $k_\mathrm{r_{max}gm}$, and $k_\mathrm{r_{max}A}$ are the maximum observed growth rates under optimum substrate temperature, feed, and air concentrations respectively.
Since the larvae also undergo morphological changes during the larval phase, referred to as instars, it is necessary to know the development age at which these transitions occur.
For this purpose, we introduced the concept of development sums $T_\mathrm{\Sigma}$ in our previous work \citep{Padmanabha2020} which is the integration of growing conditions dependent development rates that indicate how much the larvae have developed through the larval stage.
However, analysis of the experiments reveal that the final transformation to prepupae, referred to as maturity in this work, results in elevated metabolic rates.
The maturity process representing the transformation of the larvae to its final instar is now defined as a switching process that is activated when the larvae reach certain development sums ($T_\mathrm{\Sigma}$). This is given as
\begin{align}\label{eq:r_lrv_mat}
	r_\mathrm{mat} &= r_\mathrm{B_{mat}}(T_\mathrm{\Sigma}) \left( \dfrac{r_\mathrm{T}(T_\mathrm{med})}{k_\mathrm{r_{max}T}} \ \dfrac{r_\mathrm{F_{grw}}(B_\mathrm{feed})}{k_\mathrm{r_{max}gm}}  \ \dfrac{r_\mathrm{A}(A_\mathrm{air})}{k_\mathrm{r_{max}A}} \right),\\
	r_\mathrm{B_{mat}}(T_\Sigma) &= \begin{cases}
		0 		& \text{if } T_\Sigma < k_\mathrm{T_\Sigma 1} \\
		1 		& \text{if } k_\mathrm{T_\Sigma 1} < T_\Sigma < k_\mathrm{T_\Sigma 3} \\
		0 		& \text{if } T_\Sigma \geq k_\mathrm{T_\Sigma 3} \\
		\end{cases}
\end{align}
where $r_\mathrm{B_{mat}}$ is the maturity process activation function that only activates when development sums is between $k_\mathrm{T_\Sigma 1}$ and $k_\mathrm{T_\Sigma 3}$.

We also introduce a new state variable for the larvae model in this work to consider the wet body mass change.
Modelling of the wet mass change of the larvae is of interest since it allows for accurate modelling of the water flux and also implementing the in-situ mass monitoring in the production process. 
This wet mass of the larvae can be obtained as the sum of dry mass and the water balance through assimilation and maintenance as
\begin{equation}\label{eq:d_wet/dt}
	\frac{\mathrm{d}B_\mathrm{wet}}{\mathrm{d}t} = 	\frac{\mathrm{d}B_\mathrm{dry}}{\mathrm{d}t} + \phi_\mathrm{W_{assim}} - \phi_\mathrm{W_{maint}},
\end{equation}
where $\phi_\mathrm{W_{assim}}$ and $\phi_\mathrm{W_{maint}}$ are the water assimilated into the larva and spent for the metabolic activity respectively.
Finally, in this work, we consider the feed in substrate $B_\mathrm{feed} = N_\mathrm{feed}$ and substrate moisture concentration $W_\mathrm{med\%} = \left(\dfrac{W_\mathrm{med}}{W_\mathrm{med}+N_\mathrm{feed}}\right)$ and air concentration $A_\mathrm{air}=\dfrac{O_\mathrm{air}}{C_\mathrm{air}}$.
The remaining model equation corresponding to the larvae are used as presented by \cite{Padmanabha2020}, unchanged.

\subsubsection{Energy and resource flux components contributed by larvae and microbiome}
The resource fluxes (heat, gas, humidity/water) contributed by different processes i.e. assimilation respiration, maintenance, maturity and microbiome activity are modelled in this section. 

The metabolic activities are exothermic both in the larvae and the microbiome present in the substrate and thus generate heat as a byproduct.
Microbiome gets introduced into the growing medium through the feed and also the guts of the larvae. This cannot be avoided and should be considered as part of the process. 
The heat produced by the larvae and the microbiome in the growing medium can be described as a sum of different metabolic processes as 
\begin{equation}\label{eq:phi_qbio}
	\phi_\mathrm{Q_{bio}} = L_\mathrm{num} (k_\mathrm{Q_{assim}} \phi_\mathrm{B_{assim}} + k_\mathrm{Q_{maint}} \phi_\mathrm{B_{maint}} + k_\mathrm{Q_{mat}} \phi_\mathrm{B_{mat}}) + k_\mathrm{Q_{bio}} N_\mathrm{med},
\end{equation}
where $L_\mathrm{num}$ is the number of larvae in the substrate, $k_\mathrm{Q_{assim}}$, $k_\mathrm{Q_{maint}}$, and $k_\mathrm{Q_{mat}}$ represents the heat produced due to the assimilation, maintenance and maturity of the larvae respectively and $k_\mathrm{Q_{bio}}$ is the heat produced by the microbiome per gram of feed in growing medium.
Net water assimilated by the larvae is modelled to be proportional to the rate of assimilation of feed and maintenance expense as
\begin{equation}\label{eq:lrv_W_assim}
	(\phi_\mathrm{W_{assim}} - \phi_\mathrm{W_{maint}}) = \phi_\mathrm{B_{assim}} k_\mathrm{W_{assim}}  W_\mathrm{med\%} - \phi_\mathrm{B_{maint}} k_\mathrm{W_{assim}},
\end{equation}
where $k_\mathrm{w_{assim}}$ is the specific water consumption per gram feed assimilated by the larvae.
The influence of the metabolic activity on the gas flux can be modelled as
\begin{equation}
\begin{split}\label{eq:phi_cbio}
\phi_\mathrm{C_{bio}} &= L_\mathrm{num} (k_\mathrm{C_{assim}} \phi_\mathrm{B_{assim}} + k_\mathrm{C_{maint}} \phi_\mathrm{B_{maint}}+k_\mathrm{C_{mat}} \phi_\mathrm{B_{mat}}) + k_\mathrm{C_{bio}} N_\mathrm{feed} W_\mathrm{med\%}, \\
\phi_\mathrm{O_{bio}} &= k_\mathrm{bio_{C:O}} \phi_\mathrm{C_{bio}},
\end{split}
\end{equation}
where $k_\mathrm{C_{assim}}$, $k_\mathrm{C_{maint}}$, and $k_\mathrm{C_{assim}}$ are specific CO$_2$ production rates of the assimilation, maintenance and maturity process respectively, $k_\mathrm{C_{bio}}$ is the specific CO$_2$ production due to the microbiome growth and $k_\mathrm{bio_{C:O}}$ is the carbon-oxygen ratio of all the processes.
The feed in growing medium is mostly consumed by larvae and partly by microbiome.
Ignoring the feed consumption by microbiome, the feed ingestion and excretion flux from the substrate are given respectively as
\begin{align} \label{eq:phi_feed}
\phi_\mathrm{N_{ing}} &=  L_\mathrm{num}( r_\mathrm{assim} \ k_\mathrm{inges} \ B_\mathrm{dry}) \\
\phi_\mathrm{N_{exc}} &=  L_\mathrm{num}(k_\mathrm{\alpha_{excr}} r_\mathrm{assim} \ k_\mathrm{inges} \ B_\mathrm{dry}).
\end{align}
With the resource fluxes contributed by the larvae and microbiome modelled, the resource and energy fluxes in the production environment are modelled next.

\subsubsection{Modelling resource and energy dynamics in larvae production}\label{sec:me_model}
\begin{figure}[!hbt]
	\centering
	\def\svgwidth{\linewidth}
	\fontsize{6.5}{8}\selectfont
	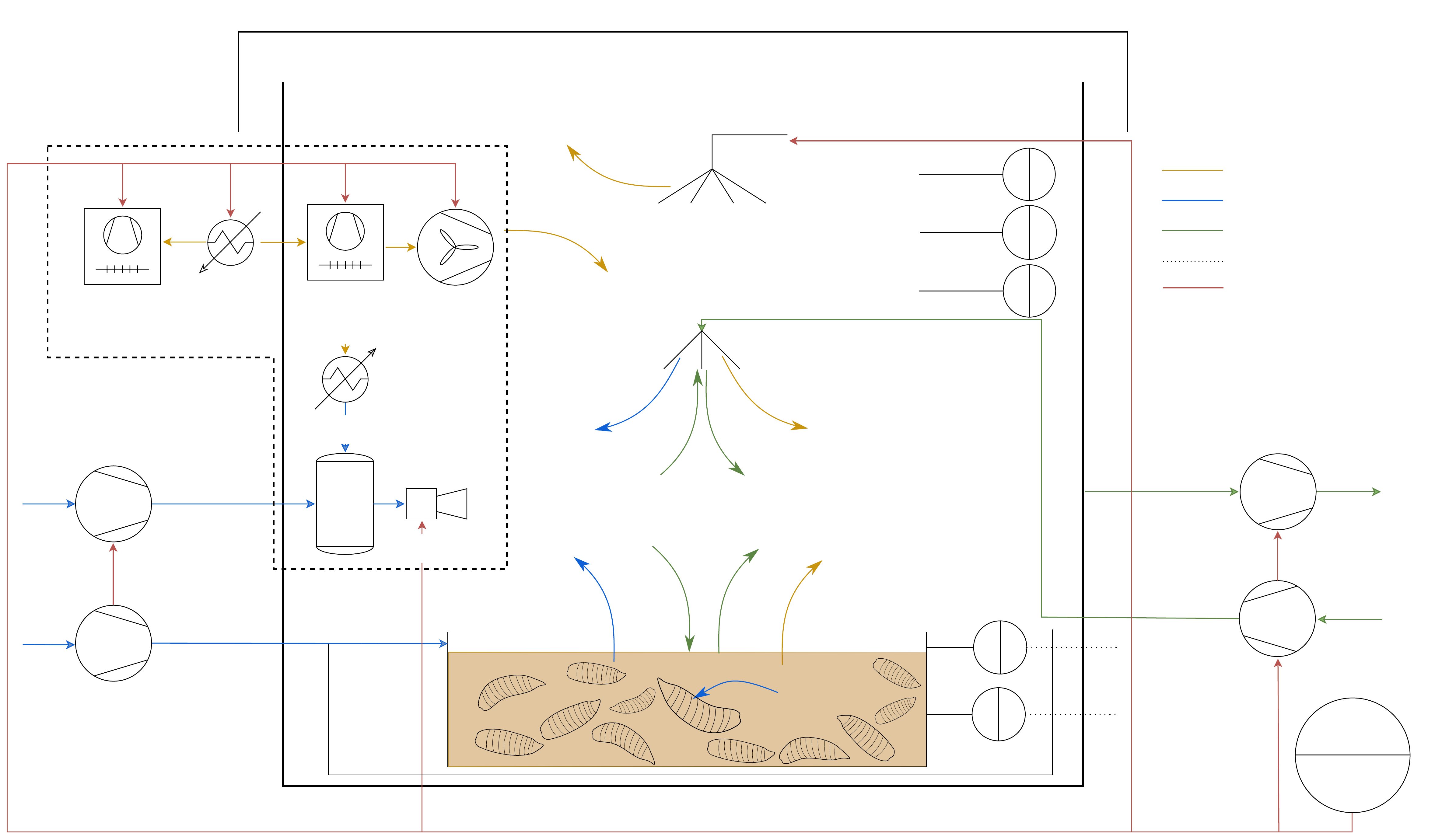\caption{{\bf Production unit components and fluxes in larvae production.} Production unit with sensors (S1-S5), air pumps (M1-M2), water pumps (M3-M4), air conditioning unit (based on thermoelectric cooler (TEC) for heating, cooling, condensation), humidifier, LED lighting as presented in \cite{Padmanabha2019}. 
	}
	\label{fig:growth_chamber}
\end{figure}

An overview of the actuators that influence the inputs, sensors that measure certain states and the important energy and mass fluxes are visualized in Fig~\ref{fig:growth_chamber}. 
This represents a fully closed or reactor type production setup presented in \cite{Padmanabha2019} and is referred to as production unit in this work.
The models proposed in this work is derived for the complex reactor type setup and identifies all the flux components.
These flux terms and the corresponding model parameter values can be adapted to different production setups by performing parameter calibration.

Mass fluxes (e.g. water, water vapor, CO$_2$, O$_2$) and energy fluxes (e.g. heat and light) in the production unit can be mechanistically described using mass and energy balance equations.
These fluxes and the influence they have on various states are presented as differential equations that describe the rate change of the states in response to the net flux changes.

Firstly, the rate of change of temperature of the air inside the production unit can be modelled as
\begin{equation}\label{eq:dT/dt}
	k_\mathrm{C_{air}}\frac{\mathrm{d}T_\mathrm{air}}{\mathrm{d}t} = \phi_{\mathrm{Q_{LED}}} + \phi_{\mathrm{Q_{hx-a}}} + \phi_{\mathrm{Q_{exch}}}+ \phi_{\mathrm{Q_{leak}}} + \phi_{\mathrm{Q_{door}}} + \phi_{\mathrm{Q_{m-a}}} + \phi_{\mathrm{Q_{{a-c}}}},
\end{equation}
where $k_\mathrm{C_{air}}$ is the total heat capacity of the air in the production unit, $\phi_{\mathrm{Q_{LED}}}$, $\phi_{\mathrm{Q_{hx-a}}}$ and $\phi_{\mathrm{Q_{exch}}}$ are the heat fluxes contributed by the actuators (LED, heater-cooler and ventilator pumps respectively), $\phi_{\mathrm{Q_{leak}}}$ and $\phi_{\mathrm{Q_{door}}}$ are the losses due to leakage and the door opening event respectively, $\phi_{\mathrm{Q_{m-a}}}$ is the convective heat flux between the air and the growing medium and $\phi_{\mathrm{Q_{{a-c}}}}$ is the convective flux between the air and the walls of the production environment.

Change in temperature in the growing medium can be summarized as
\begin{equation}\label{eq:dT_med/dt}
	k_\mathrm{C_{med}}\frac{\mathrm{d}T_\mathrm{med}}{\mathrm{d}t} = \phi_{\mathrm{Q_{bio}}} - \phi_{\mathrm{Q_{m-a}}} - \phi_{\mathrm{Q_{m-c}}} - \phi_{\mathrm{Q_{L,med}}},
\end{equation}
where $k_\mathrm{C_{med}}$ is the total heat capacity of the growing medium, $\phi_{\mathrm{Q_{bio}}}$, $\phi_{\mathrm{Q_{m-c}}}$, and $\phi_{\mathrm{Q_{L,med}}}$ are, respectively, the heat fluxes due to the metabolic activity of larvae and microbiome, conductive heat transfer to the walls and latent heat of evaporation and condensation taking place in the growing medium.

The heater-cooler system consists of a heat exchanger with mass and stores heat during operation. Therefore the heat transfer from the heat exchanger to the air depends on its current temperature which can be modelled as
\begin{equation}\label{eq:dT_hx/dt}
	k_\mathrm{C_{hx}}\frac{\mathrm{d}T_\mathrm{hx}}{\mathrm{d}t} = \phi_{\mathrm{Q_{TEC}}} - \phi_{\mathrm{Q_{hx-a}}} + \phi_{\mathrm{Q_{c-hx}}} - \phi_{\mathrm{Q_{L,hx}}},
\end{equation}
where $k_\mathrm{C_{hx}}$ is the heat capacity of the heat exchanger, $\phi_{\mathrm{Q_{TEC}}}$ is the heat energy supplied by TEC, $\phi_{\mathrm{Q_{hx-a}}}$, $\phi_{\mathrm{Q_{c-hx}}}$ and $\phi_{\mathrm{Q_{L,hx}}}$ are the convective transfer to air, conductive transfer to walls and latent heat of condensation and evaporation from the surface of the heat exchanger.

The temperature on the walls of the production unit varies depending on the air inside $T_\mathrm{air}$, the air outside $T_\mathrm{out}$, and its total heat capacity.
This heat change is given as
\begin{equation}\label{eq:dT_chm/dt}
	k_\mathrm{C_{chm}}\frac{\mathrm{d}T_\mathrm{chm}}{\mathrm{d}t} = -\phi_{\mathrm{Q_{a-c}}} - \phi_{\mathrm{Q_{c-o}}} - \phi_{\mathrm{Q_{c-hx}}} - \phi_{\mathrm{Q_{L,chm}}},
\end{equation}
where $k_\mathrm{C_{chm}}$ is the heat capacity of the walls, $\phi_{\mathrm{Q_{c-o}}}$, $\phi_{\mathrm{Q_{c-TEC}}}$ and $\phi_{\mathrm{Q_{L,chm}}}$ are the convective heat loss to the outside air, conductive heat flux from heat exchanger to the walls and the latent heat of condensation and evaporation from wall respectively.

The humidity in the production unit can be modelled as the mass transfer between different surfaces as
\begin{equation}\label{eq:dH/dt}
	k_\mathrm{V_{chm}} \frac{\mathrm{d}H_\mathrm{air}}{\mathrm{d}t}	= \phi_\mathrm{H_{u}} - \phi_\mathrm{H_{exch}} - \phi_\mathrm{H_{leak}} - \phi_\mathrm{H_{door}} - \phi_\mathrm{W_{L,chm}} - \phi_\mathrm{W_{L,hx}} + \phi_\mathrm{W_{L,med}},
\end{equation}
where $k_\mathrm{V_{chm}}$ is the inner volume of the production unit, $\phi_\mathrm{H_{u}}$, $\phi_\mathrm{H_{exch}}$, $\phi_\mathrm{H_{leak}}$, $\phi_\mathrm{H_{door}}$ are the water vapor fluxes due to the humidifier, ventilation, leakage, and door opening events respectively, $\phi_\mathrm{W_{L,chm}}$, $\phi_\mathrm{W_{L,hx}}$, and $\phi_\mathrm{W_{L,med}}$ are evaporation-condensation on the production unit walls, heat exchanger surface and the growing medium surface respectively.

Change in water quantity in the growing medium is given by
\begin{equation} \label{eq:dW_med/dt}
	\frac{\mathrm{d}W_\mathrm{med}}{\mathrm{d}t} = \phi_\mathrm{W_{u}} + \phi_\mathrm{W_{L,med}} - \phi_\mathrm{W_{bio}},
\end{equation}
where $\phi_\mathrm{W_{u}}$, $\phi_\mathrm{W_{L,med}}$, $\phi_\mathrm{W_{bio}}$ are the water fluxes due to the input water pump, evaporation-condensation from the air-growing medium surface, and the larvae and microbiome assimilation process respectively.

Due to the temperature gradients between the surfaces of the production unit inner wall, the heat exchanger, and the air, water transport takes place resulting in evaporation and condensation.
Water condenses on the surfaces when temperature of walls are colder, influencing the thermal conductivity, and evaporates when temperature of walls are higher. When sufficient water is collected, the water forms droplets and flows into the respective collection tanks.
Quantification of this condensate will further extend the model in completing the water flux components. 
These fluxes for both heat exchanger surface and the wall surface can be modelled as
\begin{equation} \label{eq:dW_chm/dt}
	\frac{\mathrm{d}W_\mathrm{chm}}{\mathrm{d}t} = \phi_\mathrm{W_{L,chm}} - \phi_\mathrm{W_{chm,out}},
\end{equation}
\begin{equation} \label{eq:dW_TEC/dt}
	\frac{\mathrm{d}W_\mathrm{hx}}{\mathrm{d}t} = \phi_\mathrm{W_{L,hx}} - \phi_\mathrm{W_{hx,out}},
\end{equation}
where $\phi_\mathrm{W_{L,chm}}$, $\phi_\mathrm{W_{L,hx}}$ are the water flux due to condensation-evaporation and $\phi_\mathrm{W_{chm,out}}$, $\phi_\mathrm{W_{TEC,out}}$ are the condensate that trickles down respectively from the surface of the walls and TEC respectively.

The gas concentration in the production unit also changes due to the larval and microbiome metabolic activities.
The CO$_2$ and similarly O$_2$ variation can be modelled as
\begin{equation}\label{eq:dC/dt}
	k_\mathrm{V_{chm}}\frac{\mathrm{d}C_\mathrm{air}}{\mathrm{d}t}= \phi_\mathrm{C_{exch}} + \phi_\mathrm{C_{leak}} + \phi_\mathrm{C_{bio}}
\end{equation}
where $\phi_\mathrm{C_{exch}}$, $\phi_\mathrm{C_{leak}}$, and $\phi_\mathrm{C_{bio}}$ are the CO$_2$ fluxes due to ventilation, leakage and the metabolic activities of the larvae and the microbiome respectively.
Similarly, the O$_2$ flux can be either derived from Equation~\cref{eq:dC/dt} assuming that the O$_2$ production is proportional to CO$_2$ as
\begin{equation}\label{eq:dO/dt}
	k_\mathrm{V_{chm}}\frac{\mathrm{d}O_\mathrm{air}}{\mathrm{d}t}= \phi_\mathrm{O_{exch}} + \phi_\mathrm{O_{leak}} - k_\mathrm{bio_{C:O}} \phi_\mathrm{C_{bio}},
\end{equation}
where $k_\mathrm{C:O}$ represents the number of O$_2$ molecules consumed for every CO$_2$ molecules produced.

The feed in the growing medium is consumed by the larvae and the microbiome and is converted into biomass.
This massflux of the substrate dry mass can be modelled as 
\begin{equation}\label{eq:dN_med/dt}
	\frac{\mathrm{d}N_\mathrm{med}}{\mathrm{d}t} = \phi_\mathrm{N_u} - \phi_\mathrm{N_{ing}} + \phi_\mathrm{N_{exc}} - \phi_\mathrm{N_{biome}},
\end{equation}
where $\phi_\mathrm{N_u}$ represents the feed input to the growing medium, $\phi_\mathrm{N_{ing}}$ represents the feed ingested by the larvae, $\phi_\mathrm{N_{exc}}$ represents the feed excreted back, and $\phi_\mathrm{N_{biome}}$ represents the feed consumed by the microbiome.

In case of the batch feeding regime, feed is introduced only at the beginning of the growth process.
In this case, the nutrition or quality of the feed goes down over time as the larvae ingests and excretes.
This is simply modelled as an integrator integrating the excreted feed in the growing medium as
\begin{equation} \label{eq:dNW_med/dt}
	\frac{\mathrm{d}N_\mathrm{exc}}{\mathrm{d}t} = \phi_\mathrm{N_{exc}}.
\end{equation}

Finally, the rate change of total mass of the growing medium which can be tracked during production using in-situ measurements can be modelled as
\begin{equation} \label{eq:dB_med/dt}
	\frac{\mathrm{d}B_\mathrm{med}}{\mathrm{d}t} =  \frac{\mathrm{d}W_\mathrm{med}}{\mathrm{d}t} + \frac{\mathrm{d}N_\mathrm{med}}{\mathrm{d}t} +  L_\mathrm{num}\frac{\mathrm{d}B_\mathrm{wet}}{\mathrm{d}t}, 
\end{equation}
where $L_\mathrm{num}$ is the total number of larvae present in the medium and $\frac{\mathrm{d}B_\mathrm{wet}}{\mathrm{d}t}$ represents the rate of change of wet mass of the larva.

The derivation of all the flux components of the production unit are presented in detail in the \ref{app:flux_model}.

 \subsection{Data, parameter estimation and model validation}\label{sec:data_tools}
The data to perform the model validation and parameter estimation are obtained through experiments executed in the production unit described by \cite{Padmanabha2019} and are presented in Fig~\ref{fig:acsd_setup} in \ref{app:exp_setup}. Two different experiments, larvae growth and thermodynamics, were performed to obtain data with and without the substrate and larvae. These two data sets enable to identify and differentiate the influence of larvae and substrate containing microbiome on the energy and resource fluxes in the production environment.

\paragraph{Thermodynamics, heat and water flux experiments (TD1--TD6)}
Tests were performed under different conditions to study the heat and humidity fluxes in the absence of any larvae and feed.
The data obtained in these experiments are mainly used to characterize and obtain model parameters of the production setup.
The details of these experiments and the resulting goodness of fit of the models with identified parameters are included in \ref{app:td_exp} and \ref{app:td_validation} respectively.


\paragraph{Larvae growth experiments (TG1--TG3)}
The influence of temperature on larval growth and the dynamic changes of resources was studied in this experiment separately under three different fixed air temperatures $T_\mathrm{air}$ (TG1: 25~\si{\celsius}, TG2: 29~\si{\celsius} and TG3: 33~\si{\celsius}).
For this study, young larvae aged 7-10 days with mean dry mass of \SI{4.28}{\milli\gram} each and dry feed consisting of one part wheat bran dry matter and three parts pig feed were obtained from Hermetia Baruth GmbH.
Substrate was prepared by mixing one part dry feed and three parts water.
About 2000 larvae and 2.0~\si{\kilogram} of substrate were introduced in a stainless steel tray achieving a substrate depth of about 3--5~\si{\centi\meter} in each experiment.
The production units were operated with fixed temperatures, periodic ventilation at a 10~\si{\minute} ON and 20~\si{\minute} OFF cycles through air pumps.
The production units were opened on a mostly daily basis to collect larvae samples (about 90 larvae) and measure growing medium mass.
The collected larvae wet mass were measured immediately and dry mass were measured after drying at \SI{70}{\celsius} for \SI{6}{\hour}.
The raw data obtained in the experiment TG2 is illustrated in the Fig~\ref{fig:T_experiments_data} in \ref{app:tg2_raw} for reference.

A summary of all the experiments, measurements taken, and the corresponding parameters estimated are provided in Table~\ref{tab:list_par_experiments}.
The constants and model parameters used in the model are listed in Table~\ref{tab:syms_list} in \ref{app:model_params}.
 \begin{table}[!ht]
		\centering
		\caption{
			{\bf Data sets for model validation and parameter estimation}}
	 	\begin{tabular}{|p{0.2\textwidth}|p{0.2\textwidth}|p{0.2\textwidth}|p{0.25\textwidth}|}
		\hline
		Experiment & Measurement type& Measurements & Parameters estimated \\
		\thickhline
		\multirow{3}{=}{Thermodynamics: TD1--TD6}&  sensors & $C_\mathrm{out}$, $H_\mathrm{out}$, $T_\mathrm{out}$, $C_\mathrm{air}$, $H_\mathrm{air}$, $O_\mathrm{air}$, $T_\mathrm{air}$, $T_\mathrm{med}$ &  \multirow{3}{=}{
		$k_\mathrm{h_{a-c}}$, $k_\mathrm{he_{a-m}}$,$k_\mathrm{hm_{a-m}}$, $k_\mathrm{h_{a-ah}}$, $k_\mathrm{h_{o-c}}$,$k_\mathrm{U_{hx-c}}$, $k_\mathrm{U_{m-c}}$, $k_\mathrm{c_{chm}}$, $k_\mathrm{h_{med}}$, $k_\mathrm{h_{chm}}$, $k_\mathrm{h_{hx}}$ } \newline\\ \cline{2-3}
		& manual (starting)& $B_\mathrm{feed}$,$W_\mathrm{med}$, $B_\mathrm{med}$ & \\
		\hline
		\multirow{3}{=}{Larvae Growth: TG1--TG3}&  sensors & $C_\mathrm{out}$, $H_\mathrm{out}$, $T_\mathrm{out}$, $C_\mathrm{air}$, $H_\mathrm{air}$, $O_\mathrm{air}$, $T_\mathrm{air}$, $T_\mathrm{med}$ &  \multirow{3}{=}{$k_\mathrm{mat}$, $k_\mathrm{maint}$, $k_\mathrm{\alpha_{excr}}$, $k_\mathrm{\alpha_{assim}}$, $k_\mathrm{Q_{assim}}$, $k_\mathrm{Q_{maint}}$, $k_\mathrm{Q_{mat}}$, $k_\mathrm{Q_{bio}}$, $k_\mathrm{W_{assim}}$, $k_\mathrm{C_{assim}}$, $k_\mathrm{C_{maint}}$, $k_\mathrm{C_{mat}}$, $k_\mathrm{C_{bio}}$, $k_\mathrm{lrv_{C:O}}$,$k_\mathrm{c_{feed}}$} \newline\\ \cline{2-3}
		& manual (daily)& $B_\mathrm{dry}$, $B_\mathrm{wet}$ $B_\mathrm{med}$& \\ \cline{2-3}
		& manual (starting)& $B_\mathrm{feed}$, $B_\mathrm{tot}$, $B_\mathrm{wet}$, $W_\mathrm{med}$, $B_\mathrm{med}$ & \\
		\hline
	\end{tabular}
 \label{tab:list_par_experiments}
\end{table}

\paragraph{Parameter estimation and model validation}
	Models presented in this work are mostly nonlinear and therefore, nonlinear least squares data fitting method is used for parameter estimation.
	The outputs (measured variables) of the non linear system $\bs{y}$ as a function of $\bs{p}$, the model parameter to be estimated, can be defined as 
	\begin{equation}\label{eq:sys_nonlin}
		\bs{y}_k = g(\bs{x}_k, \bs{u}_k, \bs{d}_k, \bs{p}),
	\end{equation}
	where $\bs{y}_k = \bs{y}(t_k)$ and similarly $\bs{x}_k$, $\bs{u}_k$, and $\bs{d}_k$.
	The parameter estimation problem is formulated as an optimization problem with the objective of finding the parameter that minimizes the sum of square of errors as
	\begin{equation}\label{eq:lsqnonlin}
	\min_{\bs{p}} \sum^{N}_{k=1} (\hat{\bs{y}}_{k} - \bs{y}_k)^2,
	\end{equation}
	where $N$ is the number of measurements and $\hat{\bs{y}}_k$ represent the measured values.
	This parameter estimation problem was implemented in MATLAB using the \texttt{fmincon} function with direct single-shooting method with \texttt{MultiStart} to explore possible solutions within the specified boundary values of parameters.
	Simulations of dynamic models were performed using the \texttt{ode15S} solver in MATLAB to solve the differential equations.
	The ODEs representing the models are stiff due to the nonlinearities in the models (humidity and temperature change) and also the combination of slow (growth and development) and fast (evaporation, convection, etc.,) changing processes.
	The coefficient of determination ($R^2$) are calculated and used as an indicator for the goodness of fit of the developed models.

\subsection{Simulation and numerical tools}\label{sec:sim_tools}
In order to evaluate the performance of the implemented controller, the partially closed production setup described by Equation~\cref{eq:prod_sys_2} and the three Scenarios (1--3) are considered as listed in the Table~\ref{tab:oc_td}.
The starting values and external environment ($T_\mathrm{out}$, $C_\mathrm{out}$, and $H_\mathrm{out}$) measurements obtained from the growth experiments TG1--TG3 are used for the controller performance study.
\begin{table}[!h]
	\caption{
		{\bf Simulation scenarios}}
	 \begin{tabular}{|m{0.1\textwidth}|m{0.15\textwidth}|m{0.22\textwidth}|m{0.4\textwidth}|}
		\hline
		Scenario & Growing conditions & Set-point operation & Optimal controller \\ \thickhline
		1 & TG2& $T_\mathrm{air} = \SI{33}{\celsius}$\newline $H_\mathrm{air} = \SI{60}{\percent}$\newline $u_\mathrm{w_{med}}=\SI{14}{\micro\gram\per\second}$ & $\alpha_1, \alpha_2, \alpha_3 = 10$\newline $\textbf{R}=\text{diag}(0.01, 0.001, 0.001, 1000)$\newline $\textbf{S}=\text{diag}(\num{0.01}, \num{0.01}, 0.01, \num{0.01})$\\\hline
		2 & TG2 with $T_\mathrm{out}$ lowered by \SI{8}{\celsius}& $T_\mathrm{air} = \SI{33}{\celsius}$ \newline $H_\mathrm{air} = \SI{60}{\percent}$\newline $u_\mathrm{w_{med}}=\SI{14}{\micro\gram\per\second}$ &same as Scenario 1, but \newline $\textbf{R}_{3,3}=0.01$ (expensive $u_\mathrm{w_{med}}$) \\\hline
		3 & TG2& $T_\mathrm{air} = \SI{33}{\celsius}$\newline$H_\mathrm{air} = H_\mathrm{out}$\newline $u_\mathrm{w_{med}}=\SI{28}{\micro\gram\per\second}$ \newline $u_\mathrm{\Delta H}=0$ (faulty/absent)& same as Scenario 1 \newline $u_\mathrm{\Delta H}=0$ (faulty/absent)\\\hline
\end{tabular}
\label{tab:oc_td}
\end{table}
To compare the performance of the model based controller, a fixed set-point controller is chosen.
In a set-point operation, a controller not containing the knowledge of the model/process operates to maintain the predetermined growing conditions.
Under these conditions, the time taken to reach the maximum larva biomass in set-point operation is determined and used as the target time of harvest $t_\mathrm{h}$ for optimal controller.
Both controllers are applied in an open-loop setup and the resulting performance of the optimal controller is compared in terms of the resource consumed to achieve the same or more biomass at the harvest time.
 
To solve the nonlinear optimal control problem numerically, the optimization problem defined by Equation~\cref{eq:oc_problem} is formulated as a multiple shooting problem and solved using CasADi's \citep{Andersson2019} MATLAB interface together with IPOPT \citep{Waetcher2006}.
The gradients of the state constraints and objective functions with respect to the decision variables are calculated internally using CasADi's algorithmic differentiation method.
The state constraints are enforced pointwise at the boundaries of the shooting in the multiple shooting problem as described in the CasADi's documentation.
A step size or sampling and control interval of $\Delta t=\SI{1}{\hour}$ is selected for the duration of  $t\in[0, 192]$ \si{\hour}. 
The ODEs are solved using the fixed-step classic Runge-Kutta (RK4) integrator supplied with CasADi. A total of 1000 steps of step size $dt=\SI{3.6}{\second}$ is considered for each control interval.
The small integration time step is necessary to solve the stiff ODEs using the explicit method.
The focus of this work is also to generate the code of the synthesized controller (executable on different computing platforms/architecture such as PLCs, etc.) and also to parallelize the execution of the OCP for faster execution.
Therefore, CasADi's RK4 implementation is selected since it executes faster compared to \texttt{ode15s} and \texttt{CVODES} solvers and also supports code generation.

\section{Results and Discussion} 
The ordinary differential equations ~\cref{eq:dB_dry/dt_new,eq:d_wet/dt,eq:dT/dt,eq:dT_med/dt,eq:dT_hx/dt,eq:dT_chm/dt,eq:dH/dt,eq:dW_med/dt,eq:dW_chm/dt,eq:dW_TEC/dt,eq:dC/dt,eq:dO/dt,eq:dN_med/dt,eq:dNW_med/dt,eq:dB_med/dt} model the mass and energy dynamics of the entire production system.
In this section, the data obtained from the experiments and the results of the parameter estimation process for the production unit model and the larvae model are discussed. 
The goodness of fit of the developed model is evaluated based on the data sets from the experiments.
The results of the optimal control studies are presented along with its performance in reducing the resource consumption. 

\subsection{Model goodness of fit for biomass, heat, water, gases, and larvae growth}\label{sec:results_model}
\begin{figure}[!h]
	\includegraphics[width=\linewidth,trim={0 10 0 40},clip]{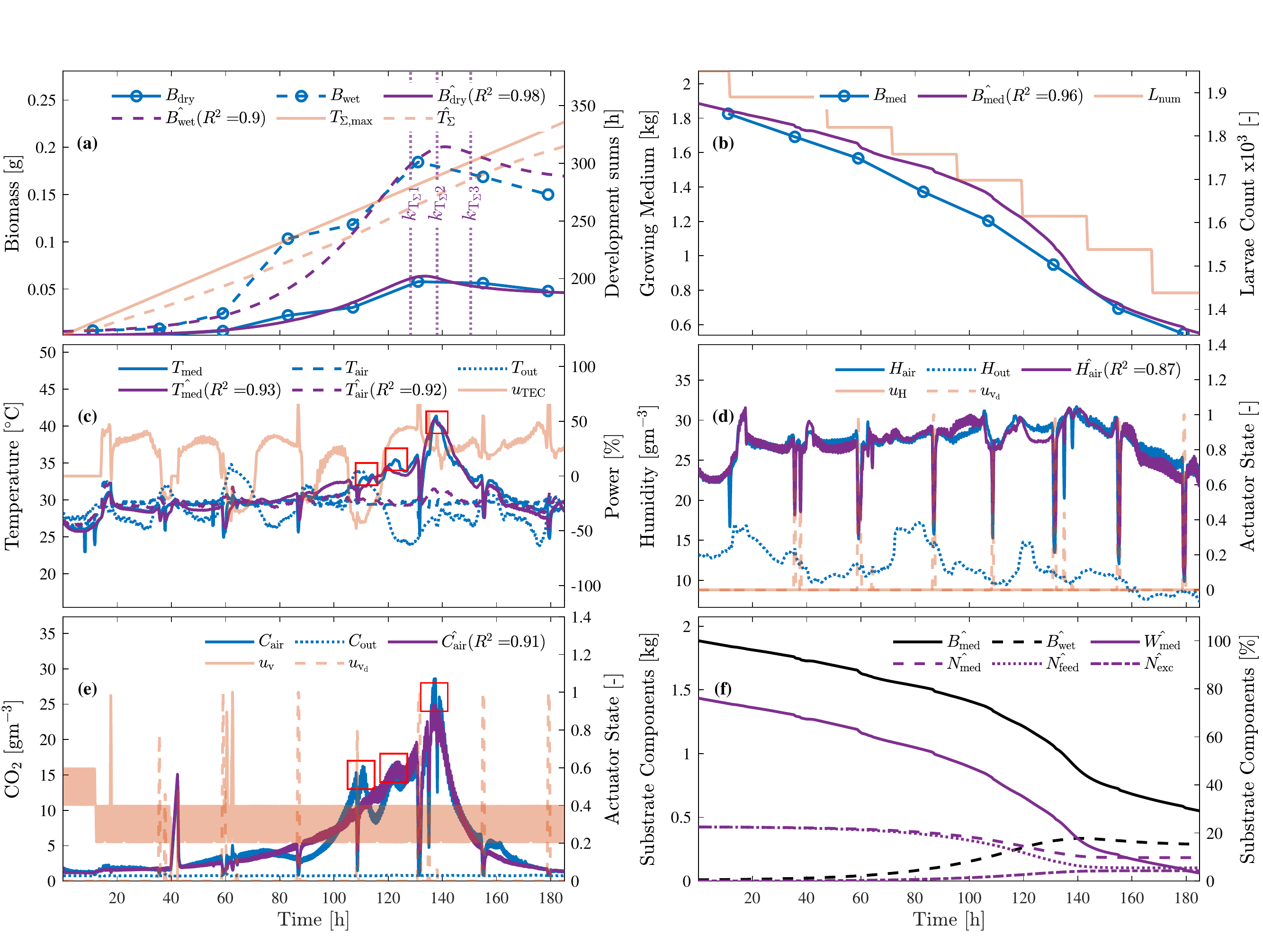}\caption{\textbf{Measurement vs. simulation (entities with hat): biomass, energy and resource changes during growth and development.}
		Data from TG2 experiment with air temperature set to a constant \SI{29}{\celsius}.
		\textbf{(a)} Wet and dry mass of individual larva and the corresponding development sums.
		\textbf{(b)} Mass change of the total growing medium including the larvae mass, feed, excreta and water.
		\textbf{(c)} Temperature changes in the growing medium and production environment.
		\textbf{(d)} Humidity changes in production environment.
		\textbf{(e)} CO$_2$ concentration changes in production environment.
		\textbf{(f)} Substrate component changes such as water, feed, excreta, and larvae wet mass as computed by the model.}
	\label{fig:lrv_model_valid}
\end{figure}

\begin{figure}[!h]
	\includegraphics[width=\linewidth,trim={20 0 10 0},clip]{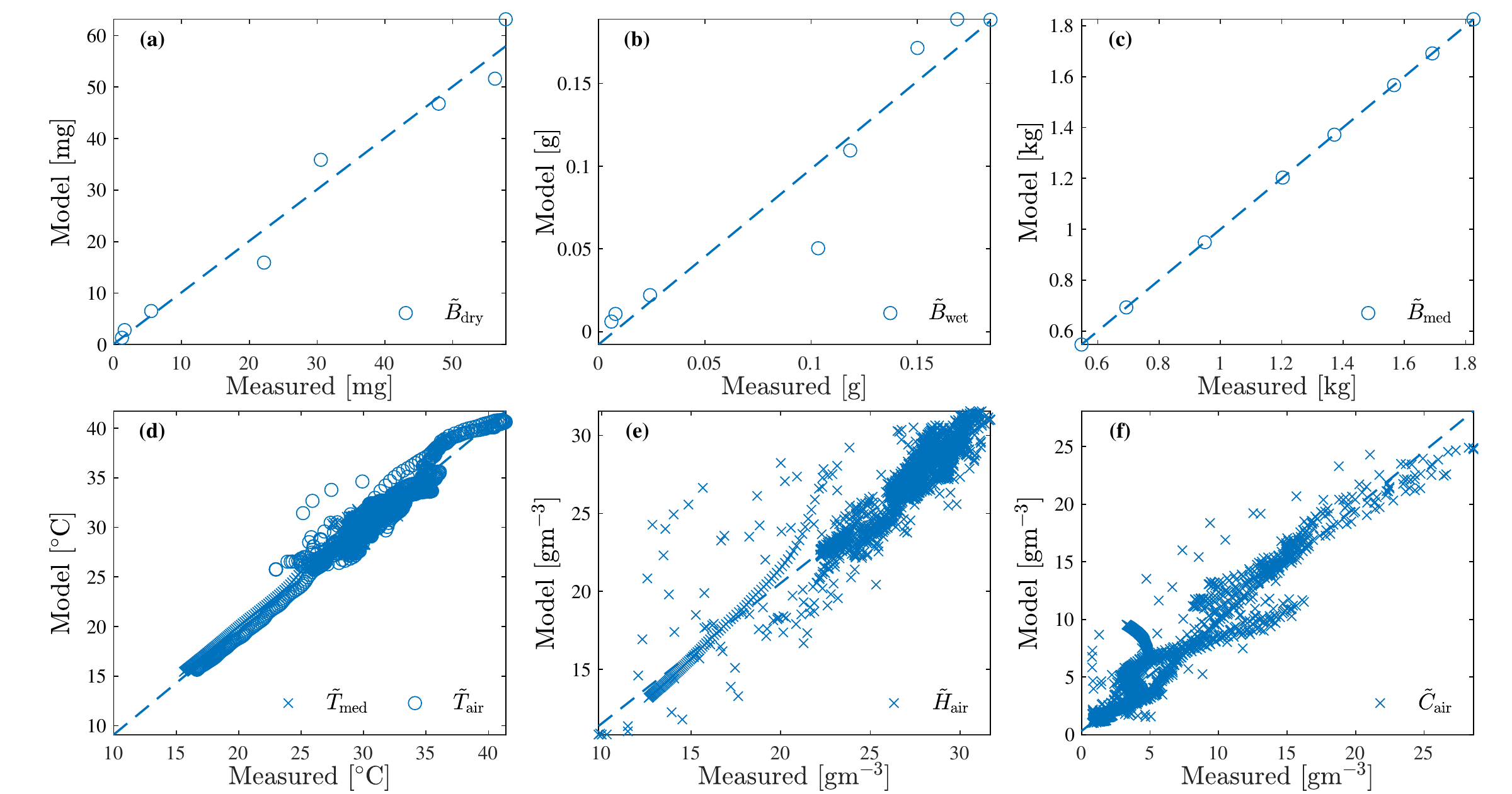}\caption{\textbf{Measurement vs. simulation: residuals analysis.}
		Measured data from TG2 experiment is plotted against the model simulation data.
		\textbf{(a)--(f)} represent the deviation of the model data from the measured data of larva dry mass, larva wet mass, growing medium mass, temperature of air, temperature of growing medium, humidity, and CO$_2$ concentration respectively.}
	\label{fig:lrv_model_residuals}
\end{figure}

This section addresses the performance of the models in describing the fluxes contributed mainly due to the presence and the growth of the larvae.
Measurements from experiment TG2, after applying moving average filter of window size 50 (for noise reduction) and resampling, are plotted and compared to the results produced by the models in Fig~\ref{fig:lrv_model_valid}. In addition, to further analyze the model quality, the deviation of the model from the measured data are presented in Fig~\ref{fig:lrv_model_residuals}.
The larvae wet mass, dry mass, and the growing medium mass were measured intermittently requiring the opening and closing of the doors.
This causes drop in the values of temperature, humidity and CO$_2$ concentration as seen in Fig~\ref{fig:lrv_model_valid}\textbf{(c)}, \textbf{(d)}, and \textbf{(e)} respectively.

Conclusion on the quality of the models in describing the variations of biomass, energy, and resources can be made based on the following observations.
Firstly, the larvae biomass changes and the development stages determined by the model reproduces the measurements with a high coefficient of determination ($R^2>0.96$) for the dry mass and slightly lower score for the wet mass (see Fig~\ref{fig:lrv_model_valid}\textbf{(a)}, Fig~\ref{fig:lrv_model_residuals}\textbf{(a)}, and Fig~\ref{fig:lrv_model_residuals}\textbf{(b)}).
Secondly, the overall substrate mass $B_\mathrm{med}$ follows the measured value with slight deviation as shown in Fig~\ref{fig:lrv_model_valid}\textbf{(b)}.
Since $B_\mathrm{med}$ is modelled based on the dynamic changes of three other state variables $B_\mathrm{tot}$, $N_\mathrm{med}$, and $W_\mathrm{med}$ which are not measured independently (see Fig~\ref{fig:lrv_model_valid}\textbf{(f)}), deviation in any one state (total larvae wet mass $B_\mathrm{tot}$) translates to the deviation in the overall mass.
Thirdly, the evaporation and thus the humidity of the air inside the production system plotted in Fig~\ref{fig:lrv_model_valid}\textbf{(d)} is captured by the model also indicating the changes due to variation in temperature and the availability of the water in the growing medium.
A slightly lower $R^2=0.83$ score for $H_\mathrm{air}$, as also seen from Fig~\ref{fig:lrv_model_residuals}\textbf{(e)}, can be reasoned by the slight deviation in temperatures $T_\mathrm{med}$ and $T_\mathrm{air}$ as seen in Fig~\ref{fig:lrv_model_valid}\textbf{(c)} and Fig~\ref{fig:lrv_model_residuals}\textbf{(d)}.

Finally, an important aspect identified in the measurement of growing medium temperature $T_\mathrm{med}$ and CO$_2$ concentration $C_\mathrm{air}$, is the presence of peaks in measurement values at \SI{110}{\hour}, \SI{120}{\hour}, and \SI{138}{\hour} marked with red squares as seen in Fig~\ref{fig:lrv_model_valid}\textbf{(c)} and \textbf{(e)}.
These peaks were observed in all the growth experiments (TG1--TG3) indicating that these peaks corresponds to the high metabolic activity of the larvae undergoing instar change.
Such observations are not previously presented in any other work and can be considered as a novel and significant contribution of this work towards better understanding of the production process.

The developed models are able to capture the final instar change (maturity phenomenon) around \SI{140}{\hour} and the resulting elevated energy ($T_\mathrm{med}=$\SI{42}{\celsius}) and resource ($C_\mathrm{air}=$\SI{26}{\gram\per\meter\cubed}) fluxes due to elevated metabolic activity.
If desired, the intermediate instar changes can also be modelled by periodically repeating the appearance of maturity fluxes $\phi_\mathrm{B_{mat}}$ in Equation~\cref{eq:dB_dry/dt_new} with the period depending on the development sums $T_\mathrm{\Sigma}$.
Since for the production process only the last instar change is of interest, the intermediate instar changes are ignored in this work.

With these observations and the obtained results, it can be concluded that the models satisfy all the model requirements from \textbf{1}--\textbf{5} and describe the process dynamics necessary for process design, automation, and resource optimization.

\subsection{Model based quantification of fluxes: observations for process design}\label{sec:results_process_analysis}
\begin{figure}[!h]
	\includegraphics[width=\linewidth,trim={0 0 0 20},clip]{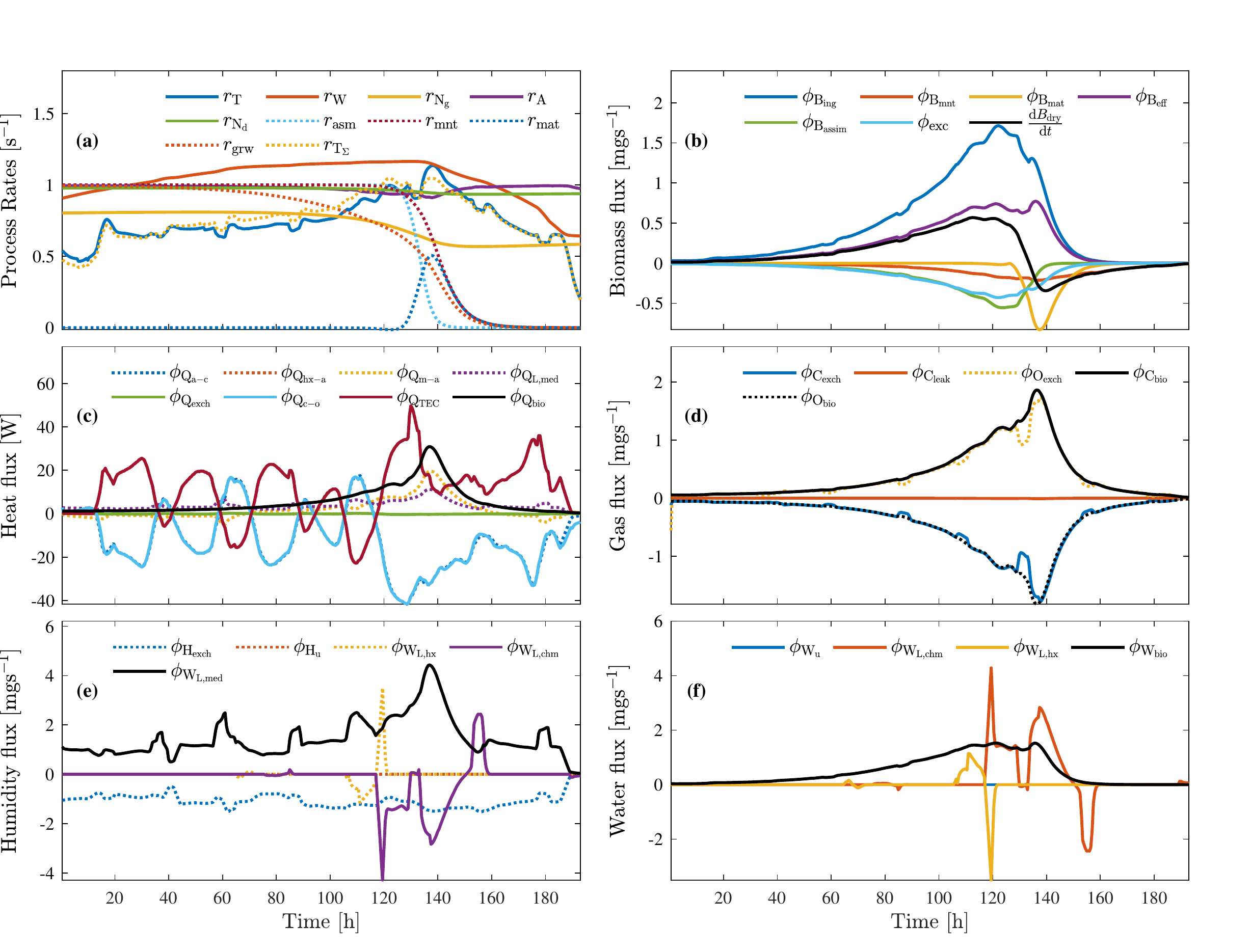}\caption{\textbf{Process rates, energy, resource, and biomass fluxes computed using the model.}
	\textbf{(a)} Process rates of all biological processes modelled for larvae growth and development
	\textbf{(b)} Biomass fluxes within the larvae.
	\textbf{(c)} Heat fluxes originating from different sources.
	\textbf{(d)} Carbon dioxide and oxygen fluxes originating due to metabolic activities and ventilation.
	\textbf{(e)} Humidity fluxes originating due to evaporation and condensation.
	\textbf{(f)} Water fluxes originating from larvae ingestion, condensation, and evaporation.}
	\label{fig:flux_model_valid}
\end{figure}

Using the models developed in this work and the identified parameters, it is possible to obtain information useful for process design.
Information such as magnitude and rate change of the individual fluxes; rate change of state variables; and influence of the state variables on the various process rates can be computed using the models.
To provide an example, all the flux terms are computed using the model and the conditions used for TG2 (initial values of the larvae number, dry mass, wet mass, feed, and water, and the climate).
These individual mass and energy flux terms and the corresponding reaction rates are presented in the Fig~\ref{fig:flux_model_valid}.

Firstly from Fig~\ref{fig:flux_model_valid}\textbf{(a)}, various process rates at all time steps can be visualized to analyze the rates which limits the growth at any given instant.
The starting values of the feed (\SI{480}{\gram} dry) and water (\SI{1.5}{\kilogram}) used in the experiment TG1--TG3 provides sufficient nutrition for the growth.
However, observing the temperature dependent growth rate $r_\mathrm{T}$, it can be stated that the low initial temperature $T_\mathrm{med}$ of \SI{27}{\celsius} results in lower growth $r_\mathrm{grw}$ and development rates $r_\mathrm{T_{\Sigma}}$.
In contrast, this temperature increases later due to metabolic activity leading to increased growth and development rates.
Using this information one can design the heating strategy where the substrate is initially heated and later turned down to operate the production process at high growth rates. 

Secondly, the initial feed and moisture in substrate can be reduced significantly at start since the larvae are small, and they cannot utilize all the available feed.
As seen in Fig~\ref{fig:flux_model_valid}\textbf{(b)}, feed gets consumed at a higher rate only when larvae reach sufficiently large size.
This strategy would reduce the feed consumed by the microbiome and the water wastage due to evaporation at the early stages.
However, this strategy may be not suitable if manual effort is involved for additional feed supply.
Design decisions regarding feeding systems and strategies can be made based on these observations.

Thirdly, heat and mass (CO$_{2}$, O$_{2}$, H$_{2}$O) fluxes plotted in Fig~\ref{fig:flux_model_valid}\textbf{(c)}--\textbf{(f)} show the dynamic variation contributed by the different (physical and biological) components during the production process.
The components, plotted in black, indicate the contribution of the fluxes from the growing medium due to the direct and indirect influence of metabolic activities.
For the chosen growing tray size of \SI{0.12}{\meter\squared}, consisting of 2000 larvae, the larvae produce \SI{31}{\watt} at the peak development phase, out of which, \SI{19}{\watt} is lost to the ambient air through convection and \SI{12}{\watt} is lost through evaporation (latent heat).
For the same setup, the CO$_2$ production and thereof O$_2$ consumption at peak development phase reaches a maximum of \SI{1.9}{\milli\gram\per\second} of which \SI{95}{\percent} is ventilated out through the ventilation system.
Similarly, the larvae consume a net \SI{1.5}{\milli\gram\per\second} water at the peak growth phase.
	The humidity flux caused by the evaporation from the growing medium reaches a maximum of \SI{4.4}{\milli\gram\per\second} when the substrate temperature $T_\mathrm{med}$ reaches the maximum value.
Due to ventilation operating at a constant rate, there exists an average vapor flux of \SI{1.3}{\milli\gram\per\second}.
Based on these values, process parameters such as actuator rates, actuator capacities, resource utilization, and resource supply rates can be computed and used to design and analyze the production systems.

Finally, another observation made using the models in this study is the optimal temperatures for growth.
From Fig~\ref{fig:lrv_model_valid}\textbf{(c)} and Fig~\ref{fig:flux_model_valid}\textbf{(c)} it can be also noted that despite the temperature inside the production reactor set to a constant \SI{29}{\celsius}, the temperature in the substrate increases accelerating the metabolic processes and thus the growth and development. 
It is therefore necessary and important to consider the substrate temperature $T_\mathrm{med}$ in lieu of the ambient temperature $T_\mathrm{air}$ to draw conclusions on the influence of the temperature on growth---contradictory to \cite{Tomberlin2009,Harnden2016,Chia2018,Shumo2019} where only $T_\mathrm{air}$ is considered. 

The above discussed applications of model for process design can be further extended for process automation and resource optimization as discussed in the next section.

\subsection{Optimization of resource and energy using model-based optimal control}\label{sec:results_oc}
The second application demonstrated in this work using the models together with the estimated parameters is the optimization of resources and energy consumption in the larvae production process.
Scenarios 1--3 defined in Section~\ref{sec:sim_tools} are executed and the performance of the synthesized controller are compared, in terms of resources consumed, with the results of set-point operation.
The first scenario compares the resource consumption between the production setups operated by the two controllers (model based controller and set-point controller) considered in nominal weather conditions and resource costs.
Scenario~2 considers extreme conditions such as colder climate condition and expensive or limited water supply. This should increase the overall operation costs in fixed/set-point operation where the process knowledge is absent.
Lastly, Scenario~3, showcases performance of the controller in operating either a further simplified setup or faulty setup with absence of some actuators for manipulating the growing conditions. In this case a humidity control device is considered to be either absent or faulty  and therefore, the process can only be controlled through temperature control and water supply to the substrate.
In all three cases it is expected that the controller with process knowledge perform better compared to the fixed set-point operation.

\begin{figure}[!h]
	\includegraphics[width=\linewidth,trim={0 0 0 25},clip]{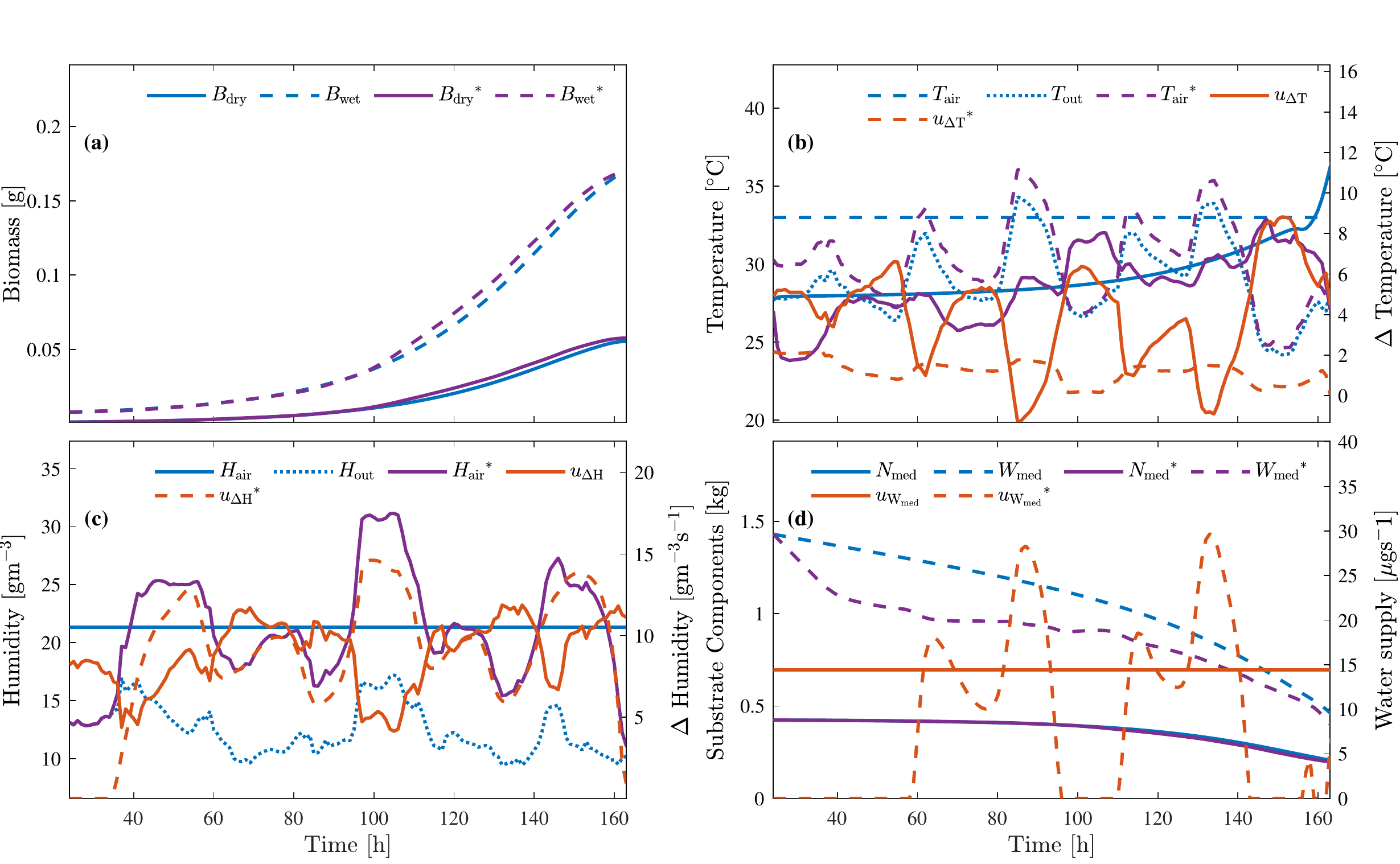}\caption{\textbf{Optimization of energy and resource consumption using optimal control (Scenario 1).} State trajectories and control inputs generated by the optimal controller are represented by $^*$.
		\textbf{(a)} Dry and wet mass of larva under constant set-points and the optimal controller case.
		\textbf{(b)} Comparison of temperature changes in substrate and air and the heat supplied.
		\textbf{(c)} Humidity modification and the additional water vapor supplied.
		\textbf{(d)} Comparison of water and feed changes in growing medium with and without optimal control.}\label{fig:oc_minimal_1}
\end{figure}

\paragraph{Scenario 1:} 
The control signals $\bs{u}^*$ generated by the synthesized controller and the resulting optimal states trajectories $\bs{x}^*$ are illustrated in Fig~\ref{fig:oc_minimal_1}.
Observing the biomass change for larva over time (see Fig~\ref{fig:oc_minimal_1}\textbf{(a)}), it can be seen that there is not a significant increase using the optimal controller.
Only a \SI{4}{\percent} increase in the biomass is noticeable.
However, a significant reduction in resource consumption is observed.
The setup with an optimal controller consumes only \SI{29}{\percent} of the energy required for heating, \SI{98}{\percent} of water and energy for humidification, and \SI{58}{\percent} of water required to keep the substrate moist.

From Fig~\ref{fig:oc_minimal_1}\textbf{(b)}, the ${u_\mathrm{\Delta T}}^*$ is high initially and gradually reduced by the controller taking into consideration the heat released due to the metabolic activities.
Similarly, from the humidification policy ${u_\mathrm{\Delta H}}^*$ as seen in Fig~\ref{fig:oc_minimal_1}\textbf{(c)}, humidity is only compensated when losses incur due to increase in temperature.
Consequently, to compensate the water loss from the substrate, water supply to growing medium only takes place when the temperature increases and the humidity drops.
Also noticeable from the results are the lower substrate moisture in the optimized case.
These policies/control trajectories generated can be further adapted adjusting the weights $\alpha$ and $\mathbf{R}$ based on the process goals.

\begin{figure}[!h]
	\includegraphics[width=\linewidth,trim={0 0 0 25},clip]{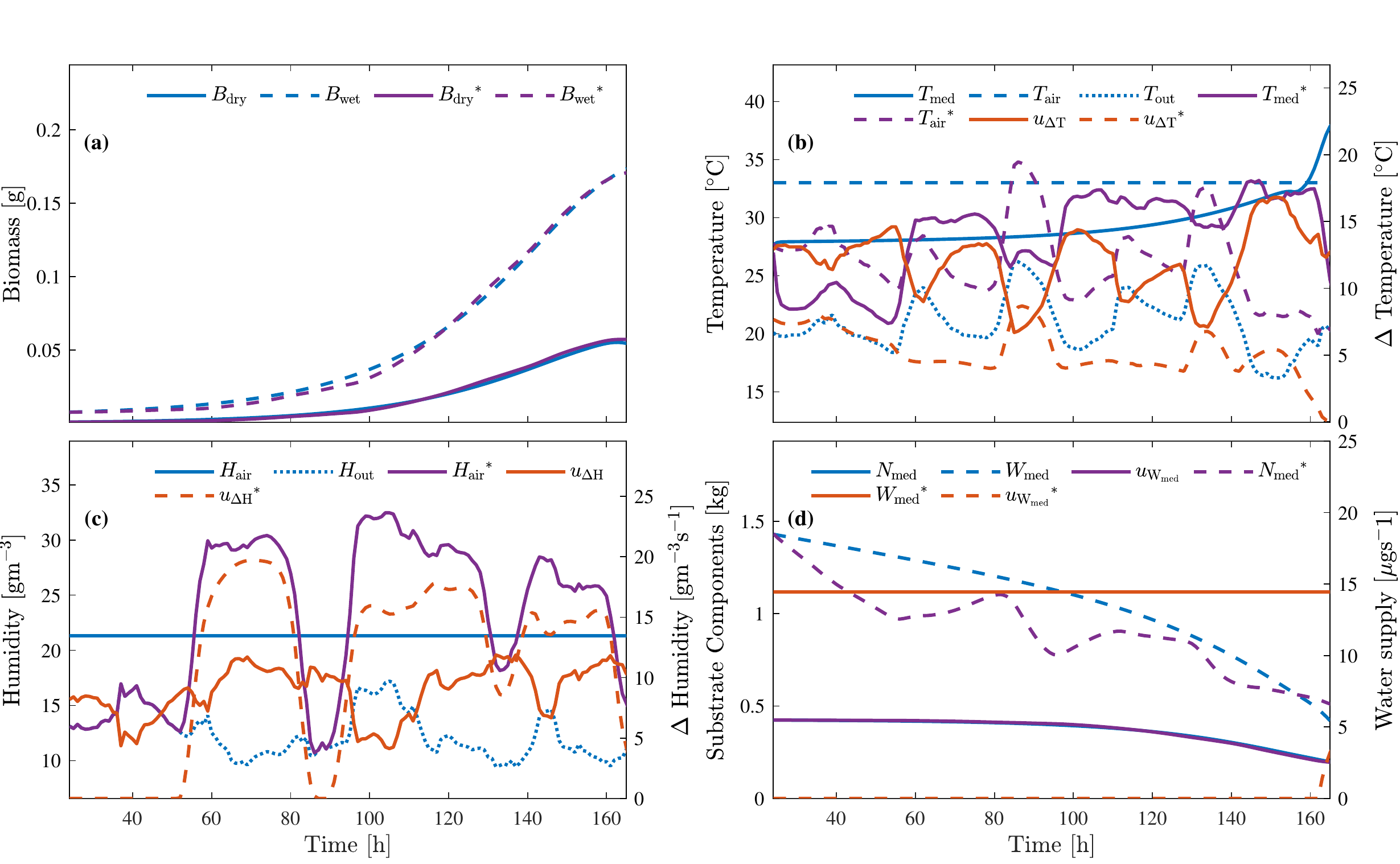}\caption{\textbf{Optimization of Energy and resource consumption using optimal control (Scenario 2).} State trajectories and control inputs generated by the optimal controller are represented by $^*$.}\label{fig:oc_minimal_2}
\end{figure}

\paragraph{Scenario 2:}
This scenario tests the operation at a outside temperature $T_\mathrm{out}$ much lower (\SI{8}{\celsius}) than considered in Scenario 1 to showcase the necessity of higher heating requirements (see Fig~\ref{fig:oc_minimal_2}(b)).
Also, the costs for water supplied to medium is set 10 times higher compared to the humidification costs.
For this scenario, the setup with an optimal controller consumes \SI{44}{\percent} of the energy required for heating, \SI{122}{\percent} of water and energy for humidification and only \SI{1}{\percent} of water required to keep the substrate moist.
There is also a small increase (\SI{3.6}{\percent}) in the biomass despite a much smaller overall energy and resource consumption.
It can be observed that the controller increases the ambient humidity to limit evaporative losses from the substrate thus requiring little to no additional supply of water to the substrate by choosing the less expensive resource.

\begin{figure}[!h]
	\includegraphics[width=\linewidth,trim={0 0 0 30},clip]{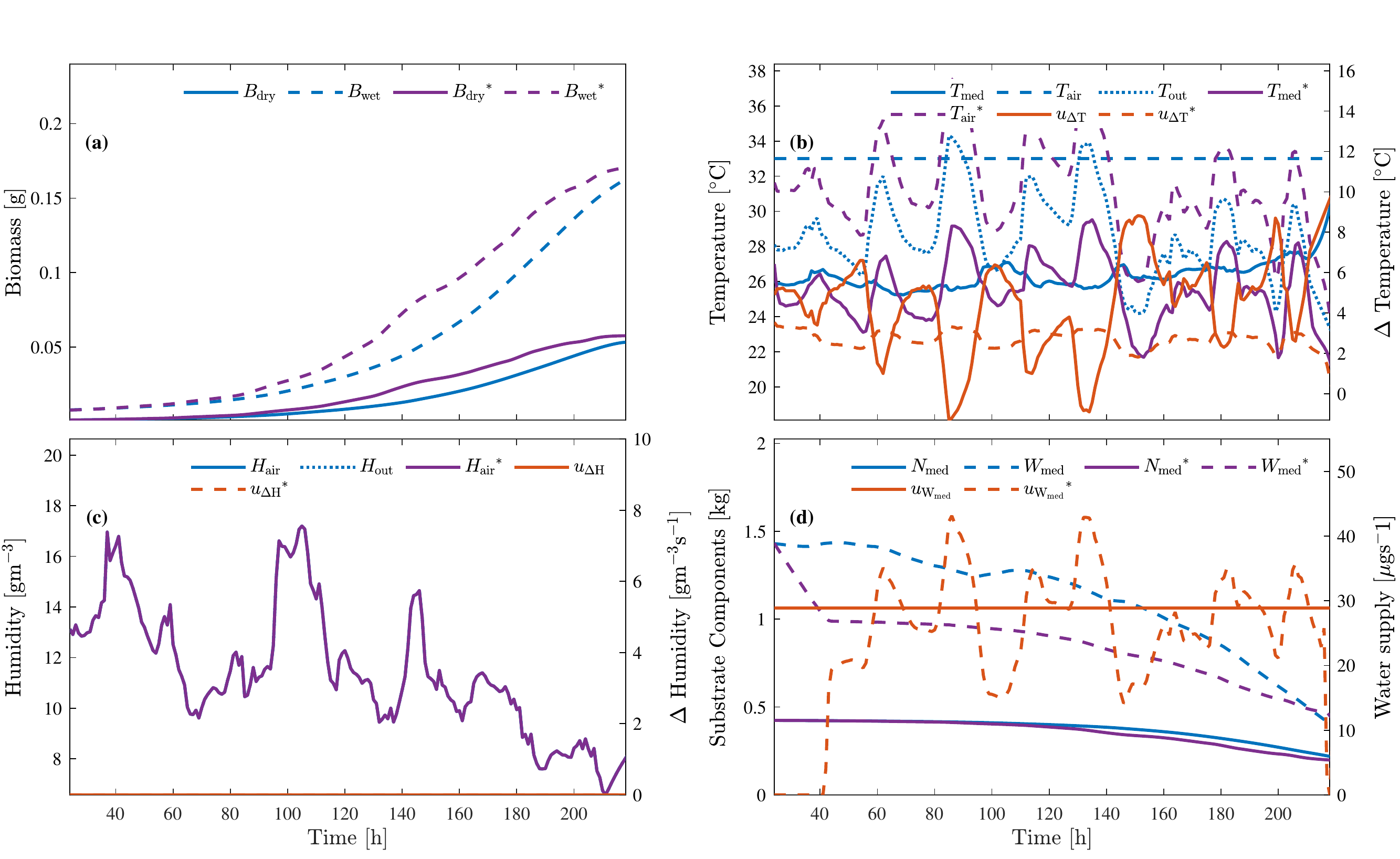}\caption{\textbf{Optimization of Energy and resource consumption using optimal control (Scenario 3).} State trajectories and control inputs generated by the optimal controller are represented by a $^*$.}\label{fig:oc_minimal_3}
\end{figure}

\paragraph{Scenario 3:}
In this scenario it is considered that no humidity control is possible and only temperature and water supply to growing medium are controlled.
For a set-point control operation, a constant higher water supply rate (\SI{28}{\micro\gram\per\second}), twice compared to scenario 1 and 2, is required to keep the substrate moist to facilitate growth for the fixed set-point operation.
In case of the optimal controller setup, only \SI{60}{\percent} of the energy is required for heating, \SI{84}{\percent} of water required to keep the substrate moist and resulting in about \SI{8}{\percent} higher biomass (see Fig~\ref{fig:oc_minimal_3}).

In all three scenarios (1-3), it was observed that the process utilizing the optimal controller generated control signals results in a significant reduction in resource consumption while still producing the same or higher larvae biomass.
It is also to be noted that the optimization is performed in these scenarios with harvest time as constant. It is possible to significantly reduce the resource consumption further by including the harvest time as an optimization variable resulting in a shorter harvesting period.
Obtaining other specific control policies are possible by either changing weights $\alpha_1$, $\alpha_2$, $\alpha_3$, $\mathbf{R}$, and $\mathbf{S}$; or modifying the objective function Equation~\cref{eq:oc_problem}; or introducing additional control variables such as harvest time and development sums for stage specific harvest.

Although the model-based optimal control approach requires detailed models with process knowledge compared to model-free or heuristic approaches, the advantages are manifold. Most importantly, an optimal controller with a process model always produces deterministic results, which allows for a sound interpretation of even those that appear to be a counter intuitive decision.
Besides that, there are several benefits in adopting the model-based optimal control framework as seen in other areas of food and bio mass production \citep{Straten2000,Padmanabha2020a,Cacho1991,Karimanzira2017,Freitas2017,Chahid2021}.
However, this is the first work that implements the model-based control and highlight its benefits for the industrial mass production of the \textit{Hermetia illucens} larvae. 


\section{Conclusion and Outlook}


Differential equations representing the dynamic changes of the resources (feed, water, larvae biomass, gas concentration, humidity, etc.) and energy in a closed production system including the interaction dynamics between the \textit{Hermetia illucens} larvae and its environment are derived based on requirements established for process automation. The derived model of the production environment together with the identified model parameters validated against the datasets 
    resulted in a high goodness of fit for all data sets. 
Furthermore, important and unique observations on the elevated heat and CO$_2$ production due to instar changes in the experiment datasets could be captured by the models.
Also, the reproduction of the elevated metabolic peaks using the newly derived model Equation~\cref{eq:dB_dry/dt_new} is also highlighted.
The differential equations representing the model~\cref{eq:dB_dry/dt_new,eq:d_wet/dt,eq:dT/dt,eq:dT_med/dt,eq:dT_hx/dt,eq:dT_chm/dt,eq:dH/dt,eq:dW_med/dt,eq:dW_chm/dt,eq:dW_TEC/dt,eq:dC/dt,eq:dO/dt,eq:dN_med/dt,eq:dNW_med/dt,eq:dB_med/dt} were validated and it is shown that they satisfy all the model requirements (1--6) defined in this work for process automation and application in optimal control.

To showcase one of the several possible use cases of the models for process design, mass fluxes, energy fluxes, and process rates were generated using the flux terms and analyzed to obtain crucial information pertaining to the production process.
We also conclude that the temperature of the growing medium is rather more important than the air temperature for the control of the larave growth and development.
In addition, conclusions for designing the feeding strategies and also designing the actuator framework could be drawn by studying the feed dynamics and fluxes of temperature, humidity, and CO$_2$ obtained using these models.
Therefore, it can be concluded that these models are novel and also very useful in designing production systems including special closed agricultural systems such as CUBEScircles \cite{Circle2018} that aim at reusing the byproducts to produce zero waste.
Not only are these models useful in designing the process, but also necessary to automate the process and simultaneously optimize the resource consumption as demonstrated using the proposed optimal control framework.

The performance of the synthesized optimal controller studied under the three scenarios, showcases its potential in optimizing resources while achieving the defined automation goals of \textit{Hermetia illucens} production.
In all three scenarios presented in this work, the optimal controller driven production process achieves same or higher larvae biomass while consuming significantly low resources compared to a fixed set-point operation. This showcases the strengths of the model based control approach which exploits the process knowledge for optimum operation.
Although its adoption is seen in other areas of food and bio mass production, this is the only work that addresses its adoption for \textit{Hermetia illucens} larvae production and presents all the necessary tools such as models (Equation~\cref{eq:dB_dry/dt_new,eq:d_wet/dt,eq:dT/dt,eq:dT_med/dt,eq:dT_hx/dt,eq:dT_chm/dt,eq:dH/dt,eq:dW_med/dt,eq:dW_chm/dt,eq:dW_TEC/dt,eq:dC/dt,eq:dO/dt,eq:dN_med/dt,eq:dNW_med/dt,eq:dB_med/dt}), data (TD1--TD6, TG1--TG3), examples (Scenario 1--3) and an objective function Equation~\cref{eq:oc_problem} to solve the control challenges associated.

The tools presented in this work will lay the foundation for future work which will focus on exploring the variations of the objective functions; adjusting the weights for fine-tuning the process for multiple process goals; define sophisticated control policies; and develop resource aware open loop decision support infrastructure and closed loop control for safe, uninterrupted, and efficient operation of closed and interconnected production systems.
\appendix
\setcounter{figure}{0}
\renewcommand\thefigure{\Alph{section}.\arabic{figure}}
\setcounter{table}{0}
\renewcommand\thetable{\Alph{section}.\arabic{table}}

\section{Flux components} \label{app:Model_flux}
Additional information pertaining to all the individual flux components and modelling of the flux components contributed due to the physical phenomenon are presented in this section.
{
	\begin{longtable}{p{0.1\linewidth}p{0.8\linewidth}p{0.15\linewidth}}
		\caption{\bf List of symbols used to describe the fluxes.} \label{tab:fluxes} \\ 
		\hline	\bf	Symbol                     & \bf Description                                		& \bf Unit                  \\ \hline \endfirsthead
		\caption[]{(continued ...)} \\
		\hline	\bf	Symbol                     & \bf Description                                		& \bf Unit                  \\ \hline 	\endhead
		\multicolumn{3}{r}{{continued on next page ...}} \\ \hline	\endfoot
		\hline \hline \endlastfoot
		$\phi_\mathrm{B_{ing}}$  & flux of feed from substrate into the larva 	& [\si{\gram\per\second}]    \\
		$\phi_\mathrm{B_{excr}}$ & flux of non digested feed back to substrate 	& [\si{\gram\per\second}]    \\
		$\phi_\mathrm{B_{assim}}$  & feed converted into energy and spent to digest the ingested feed  	& [\si{\gram\per\second}]    \\        
		$\phi_\mathrm{B_{eff}}$  & effective assimilates available from the ingested feed for growth and maintenance 	& [\si{\gram\per\second}]    \\
        $\phi_\mathrm{B_{mat}}$  & assimilates spent towards building of new structure 	& [\si{\gram\per\second}]    \\
        $\phi_\mathrm{B_{maint}}$  & assimilates spent for maintenance of existing structure 	& [\si{\gram\per\second}]    \\
		$\phi_\mathrm{W_{assim}}$    & flux of water from substrate into the larva 	& [\si{\gram\per\second}]    \\
		$\phi_\mathrm{W_{maint}}$  & water spent for maintenance respiration & [\si{\gram\per\second}]    \\
		$\phi_\mathrm{Q_{bio}}$    & heat production in growing medium & [\si{\joule\per\second}] \\
        $\phi_\mathrm{C_{bio}}$    & CO\textsubscript{2} production in growing medium & [\si{\kilogram\per\second}] \\
		$\phi_\mathrm{O_{bio}}$    & O\textsubscript{2} consumption in growing medium & [\si{\kilogram\per\second}] \\
		$\phi_\mathrm{N_{ing}}$  & net feed consumption by all larvae & [\si{\gram\per\second}] \\
		$\phi_\mathrm{N_{exc}}$  & net digested feed excretion by all larvae & [\si{\gram\per\second}] \\
		$\phi_\mathrm{N_{biome}}$  & feed consumed by microbiome & [\si{\gram\per\second}] \\
		
		$\phi_{\mathrm{Q_{LED}}}$  & heat produced by LED lighting system         & [\si{\joule\per\second}]    \\
		$\phi_{\mathrm{Q_{hx-a}}}$ & convective flux between air and heat exchanger         & [\si{\joule\per\second}]    \\
		$\phi_{\mathrm{Q_{exch}}}$ & heat flux through ventilation pumps         & [\si{\joule\per\second}]    \\
		$\phi_{\mathrm{Q_{leak}}}$ & heat flux through leakage         & [\si{\joule\per\second}]    \\
		$\phi_{\mathrm{Q_{leak}}}$ & heat flux through door         & [\si{\joule\per\second}]    \\
		$\phi_{\mathrm{Q_{m-a}}}$ & convective heat flux between growing medium and air         & [\si{\joule\per\second}]    \\
		$\phi_{\mathrm{Q_{a-c}}}$ & convective heat flux between air and walls of production system & [\si{\joule\per\second}]    \\
		$\phi_{\mathrm{Q_{m-c}}}$ & conductive heat flux between growing medium and walls of production system  & [\si{\joule\per\second}]    \\
		$\phi_{\mathrm{Q_{TEC}}}$ & heat generated or removed by the heating-cooling system & [\si{\joule\per\second}]    \\
		$\phi_{\mathrm{Q_{c-hx}}}$ & conductive heat flux between heat exchanger and walls of production system & [\si{\joule\per\second}]    \\		
		$\phi_{\mathrm{Q_{c-o}}}$ & convective heat flux between outside walls of production system and external environment & [\si{\joule\per\second}]    \\
		$\phi_{\mathrm{Q_{L,med}}}$ & heat flux in growing medium due to latent heat of evaporation and condensation & [\si{\joule\per\second}]    \\
		$\phi_{\mathrm{Q_{L,hx}}}$ & heat flux in heat exchanger due to latent heat of evaporation and condensation & [\si{\joule\per\second}]    \\
		$\phi_{\mathrm{Q_{L,chm}}}$ & heat flux in walls of production system due to latent heat of evaporation and condensation & [\si{\joule\per\second}]    \\
		$\phi_\mathrm{H_{u}}$	   & water vapor provided to or removed from the production system through (de)humidifier & [\si{\kilogram\per\second}]\\
		$\phi_\mathrm{H_{exch}}$   & water vapor change due to air exchange during ventilation & [\si{\kilogram\per\second}] \\
		$\phi_\mathrm{H_{leak}}$   & water vapor loss due to air exchange through leakage & [\si{\kilogram\per\second}] \\
		$\phi_\mathrm{H_{door}}$   & water vapor change due to air exchange during door opening events & [\si{\kilogram\per\second}] \\
		$\phi_\mathrm{W_{L,chm}}$	   & net water movement between the air and walls (condensation and evaporation) & [\si{\kilogram\per\second}]\\
		$\phi_\mathrm{W_{L,hx}}$	   & net water movement between the air and heat exchanger (condensation and evaporation) & [\si{\kilogram\per\second}]\\
		$\phi_\mathrm{W_{L,med}}$	   & net water movement between the air and growing medium (condensation and evaporation) & [\si{\kilogram\per\second}]\\
		$\phi_\mathrm{W_{u}}$	   & water provided to the growing medium through pumps & [\si{\kilogram\per\second}]\\
		$\phi_\mathrm{W_{bio}}$	   & water consumed by the larvae from the growing medium & [\si{\kilogram\per\second}]\\
		$\phi_\mathrm{W_{chm,out}}$	   & condensate flow from the walls of the production system to collection tank& [\si{\kilogram\per\second}]\\
		$\phi_\mathrm{W_{hx,out}}$	   & condensate flow from the heat exchanger to collection tank& [\si{\kilogram\per\second}]\\
        $\phi_\mathrm{C_{exch}}$   & active exchange of CO\textsubscript{2} through pumps & [\si{\kilogram\per\second}] \\		
        $\phi_\mathrm{C_{leak}}$   & passive exchange of CO\textsubscript{2} through leakage & [\si{\kilogram\per\second}] \\
        $\phi_\mathrm{O_{exch}}$   & active exchange of O\textsubscript{2} through pumps & [\si{\kilogram\per\second}] \\		
        $\phi_\mathrm{O_{leak}}$   & passive exchange of O\textsubscript{2} through leakage & [\si{\kilogram\per\second}] \\	
		$\phi_\mathrm{N_{u}}$      & feed supplied to substrate as input & [\si{\kilogram\per\second}] \\
		$\phi_\mathrm{\dot{V}_u}$  & air exchange taking place through ventilator pumps      & [\si{\meter\cubed\per\second}] \\        
	\end{longtable}
}

\subsection{Energy and resource flux components}\label{app:flux_model}
All the flux terms introduced in the mass and energy balance equations in Section~\ref{sec:me_model} are derived based on laws of physics as presented in detail here.

\paragraph{Temperature and heat fluxes:}
The individual heat flux components represented by $\phi_{\mathrm{Q}}$ in Equation~\cref{eq:dT/dt,eq:dT_chm/dt,eq:dT_hx/dt,eq:dT_med/dt} are mostly results of convection, conduction, mass transfer, and latent heat of condensation and evaporation.

Firstly, the general convective heat flux between the air and the surface of consideration is given as
\begin{equation}\label{eq:q_conv}
\begin{aligned}
	\phi_\mathrm{Q_{a-c}} 	&= k_\mathrm{A_{c}} k_\mathrm{h_{a-c}}	(T_\mathrm{chm} - T_\mathrm{air}), \quad
	\phi_\mathrm{Q_{a-m}} 	= k_\mathrm{A_{m}} k_\mathrm{h_{a-m}}	(T_\mathrm{med} - T_\mathrm{air}), \\
	\phi_\mathrm{Q_{a-hx}} 	&= k_\mathrm{A_{hx}} k_\mathrm{h_{a-hx}}	(T_\mathrm{hx} - T_\mathrm{air}), \quad
	\phi_\mathrm{Q_{c-o}} 	= k_\mathrm{A_{c}} k_\mathrm{h_{o-c}}	(T_\mathrm{out} - T_\mathrm{chm}),
\end{aligned}
\end{equation}
where the area of the surface $\mathrm{y}$ under consideration is $k_\mathrm{A_{y}}$ and the heat transfer coefficient between the air and the surface $\mathrm{y}$ is $k_\mathrm{h_{a-y}}$, and the subscripts $a$,$c$,$m$,$hx$, and $o$ corresponds to air inside, walls, growing medium, heat exchanger and air outside respectively.
The above equation for convective fluxes holds good when the ratio of mass and energy transfer is constant. 
In case of high evaporation and thus higher mass transfer, the net convective heat transfer coefficient varies proportional to the mass transfer flux \citep{Lewis1922}.
Therefore, the convective heat transfer between the air and the growing medium is modified as
\begin{equation}
	k_\mathrm{h_{a-m}} = k_\mathrm{he_{a-m}} + k_\mathrm{hm_{a-m}} \phi_\mathrm{W_{evap}}	
\end{equation}
where $k_\mathrm{he_{a-m}}$ is the convective heat transfer only due to the energy transport and $k_\mathrm{hm_{a-m}}$ is the specific mass dependent heat transfer rate.

The heat flux components due to conductive heat transfer is given as
\begin{equation}\label{eq:q_cond}
	\phi_\mathrm{Q_{c-hx}} 	= k_\mathrm{A_{hx-c}} k_\mathrm{U_{hx-c}}	(T_\mathrm{hx} - T_\mathrm{chm}), \quad
	\phi_\mathrm{Q_{m-c}} 	= k_\mathrm{A_{m-c}} k_\mathrm{U_{m-c}}	(T_\mathrm{med} - T_\mathrm{chm}),
\end{equation}
where the area of contact between the surface $\mathrm{y}$ and walls are given as $k_\mathrm{A_{y-c}}$ and the conductive heat transfer coefficient between them are given as $k_\mathrm{U_{y-c}}$ with the subscript $m$ and $hx$ corresponding to the growing medium and heat exchanger respectively.

 Heat exchange due to mass transfer with the outside environment through ventilation systems (air-pumps), leakage, and opening of door is a function of the mass transfer rate modelled as
 \begin{equation}\label{eq:q_exc}
\begin{aligned}
\phi_\mathrm{Q_{exch}} 	&= k_\mathrm{c_{air}} k_\mathrm{\rho_{air}} \overbrace{k_\mathrm{\dot{V}_{u}} u_\mathrm{V}}^{\phi_\mathrm{\dot{V}_u}}	(T_\mathrm{out} - T_\mathrm{air}), \\
\phi_\mathrm{Q_{leak}} 	&= k_\mathrm{c_{air}} k_\mathrm{\rho_{air}} k_\mathrm{\dot{V}_{leak}} (T_\mathrm{out} - T_\mathrm{air}), \\
\phi_\mathrm{Q_{door}} 	&= k_\mathrm{c_{air}} k_\mathrm{\rho_{air}} u_\mathrm{door} k_\mathrm{\dot{V}_{door}} (T_\mathrm{out} - T_\mathrm{air}),
\end{aligned}
\end{equation}
where $k_\mathrm{c_{air}}$ and	$k_\mathrm{\rho_{air}}$ are the specific heat capacity and density of the air and $k_\mathrm{\dot{V}_{u}}$, $k_\mathrm{\dot{V}_{leak}}$ and 
$k_\mathrm{\dot{V}_{door}}$ represent the maximum mass transfer rate of ventilation system, leakage, and door opening events respectively.

The heat inside the production unit is supplied and removed by the TEC based heating-cooling system.
This heat flux term contributed by the TEC module can be modelled as presented in \cite{Rowe1995}
\begin{equation}\label{eq:Q_TEC}
\phi_{\mathrm{Q_{TEC}}}= \frac{k_\mathrm{\alpha,q}u_\mathrm{T} k_\mathrm{V_{max}}}{k_\mathrm{R,q}} T_\mathrm{air} + \frac{u_\mathrm{T} {k_\mathrm{V_{max}}}^2}{2 k_\mathrm{R,q}} + k_\mathrm{TEC}(T_\mathrm{out}-T_\mathrm{air}),
\end{equation}
where $k_\mathrm{\alpha,q}$, $k_\mathrm{R,q}$, and $k_\mathrm{TEC}$ are the Seebeck coefficient, series resistance and thermal conductivity of the TEC module respectively, $u_\mathrm{T}$ is the input to the TEC driver and $k_\mathrm{V_{max}}$ is the maximum voltage that can be applied. This flux component can be replaced by the corresponding model of the heat-exchanging system in use.

The LED panel inside the production unit also produces heat and can be modeled as
\begin{equation}\label{eq:Q_led}
\phi_{\mathrm{Q_{LED}}}= \sum_{i=1}^{4}k_{\mathrm{heat}i}u_{\mathrm{I}i},
\end{equation}
where $i$ represents the narrow and wide-band wavelengths (LED channels) supported by the light panel and $k_{\mathrm{heat}i}$ is the maximum heat dissipated by the respective LED channel.


The growing medium containing high concentration of water contributes to evaporation resulting in the loss of heat through latent heat of evaporation.
This can be described as the heat required to raise the water at temperature $T_\mathrm{med}$ to \SI{100}{\celsius} plus the additional latent heat for converting water from liquid to vapor state as
 \begin{equation}\label{eq:q_evap}
\phi_\mathrm{Q_{L,med}}  	= \left(k_\mathrm{c_{water}} \left(100 - T_\mathrm{med}\right) + k_\mathrm{h_{ew}}\right) \phi_{\mathrm{W_{L,med}}},
 \end{equation}
where $k_\mathrm{h_{ew}}$ latent heat of vaporization of water at \SI{100}{\celsius}, $k_\mathrm{C_{water}}$ is the specific heat capacity of water, and $\phi_{\mathrm{W_{L,med}}}$ is the rate of transport of water between the air and the surface of the medium. 
Equation~\cref{eq:q_evap} can be applied to obtain the latent heat for walls $\phi_\mathrm{Q_{L,chm}}$ and heat exchanger $\phi_\mathrm{Q_{L,hx}}$ surface by replacing the temperature $T_\mathrm{med}$ with $T_\mathrm{chm}$ and $T_\mathrm{hx}$ respectively.

The heat capacity $k_\mathrm{C_{air}}$ of the air in production environment and $k_\mathrm{C_{med}}$ for the growing medium varies due to changes due to mass fluxes.
Therefore, these parameters are not constants but are dependent on the state variables given as
\begin{equation}
\begin{aligned}
k_\mathrm{C_{med}} &= k_\mathrm{c_{tray}} k_\mathrm{m_{tray}} + k_\mathrm{c_{feed}} B_\mathrm{feed} + k_\mathrm{c_{water}} W_\mathrm{med}\\
k_\mathrm{C_{air}} &= k_\mathrm{c_{air}} k_\mathrm{\rho_{air}} k_\mathrm{V_{chm}} + k_\mathrm{c_{vap}} k_\mathrm{V_{chm}} H_\mathrm{air},\\
k_\mathrm{C_{chm}} &= k_\mathrm{c_{chm}} k_\mathrm{m_{chm}} + k_\mathrm{c_{water}} W_\mathrm{cham},
\end{aligned}
\end{equation}
where $k_\mathrm{c_{tray}}$, $k_\mathrm{c_{feed}}$, $k_\mathrm{c_{water}}$, $k_\mathrm{c_{air}}$, $k_\mathrm{c_{vap}}$ are the specific heat capacities of growing tray, feed, water, production unit internal parts, air, and water vapor respectively, $k_\mathrm{V_{chm}}$ is the volume inside the production unit, $k_\mathrm{m_{tray}}$, and $k_\mathrm{m_{chm}}$ are the mass of the growing trey and the inner components and walls of production unit.

\paragraph{Water and humidity fluxes:}
Water fluxes in the production unit take place in both liquid and vapor forms.
Evaporation of water from the growing medium has twofold effect: cooling of the growing medium resulting in lower temperatures for larval growth and loss of moisture resulting in reduced ingestion rate.
It is therefore important to accurately model the phenomenon that captures both water loss and temperature drop.
The maximum rate of evaporation or condensation from any surface is defined as \citep{Monteith1981},
\begin{equation} \label{eq:gen_evap_flux}
	\phi_\mathrm{W_{L}} =  k_\mathrm{A_{surf}} k_\mathrm{h_{surf}} \left(H_\mathrm{sat}\left(T_\mathrm{surf}\right)-H_\mathrm{surf}\right),
\end{equation}
where $k_\mathrm{A_{surf}}$, $k_\mathrm{h_{surf}}$, and $T_\mathrm{surf}$ are the area, vapor transport coefficient and temperature of the given surface.
The saturation concentration of water vapor, $H_\mathrm{sat}$, for any given reference temperature $T_\mathrm{ref}$, can be calculated using the Magnus-Tetens equation \citep{Murray1967,Alduchov1996} as
\begin{equation}
H_\mathrm{sat}(T_\mathrm{ref}) = \frac{k_\mathrm{W_{mmol}}}{k_\mathrm{R_g}(T_\mathrm{ref}+273)}
\left(0.61094\cdot \exp\left(\dfrac{17.625\cdot T_\mathrm{ref}}{T_\mathrm{ref}+243.03}\right)\right),
\end{equation}
where $k_\mathrm{W_{mmol}}$ is molar mass of water and $k_\mathrm{R_g}$ is the gas constant.

In order to obtain actual condensation and evaporation rate from the above Equation~\cref{eq:gen_evap_flux} as separate flux terms, it is necessary to split the components and consider the actual quantity of water available for this process.
\begin{equation}\label{eq:gen_evap_cond}
\begin{aligned}
	\phi_\mathrm{W_{evap}} &=  k_\mathrm{A_{surf}} k_\mathrm{h_{surf}} \epsilon_\mathrm{evap} \max\left(H_\mathrm{sat}\left(T_\mathrm{surf}\right)-H_\mathrm{surf}, 0\right),\\
	\phi_\mathrm{W_{cond}} &=  k_\mathrm{A_{surf}} k_\mathrm{h_{surf}} \max\left(H_\mathrm{surf} - H_\mathrm{sat}\left(T_\mathrm{surf}\right),0\right),\\
	\phi_\mathrm{W_{L}} &= \phi_\mathrm{W_{evap}} - \phi_\mathrm{W_{cond}},
\end{aligned}
\end{equation}
where $\epsilon_\mathrm{evap}$ is the evaporation coefficient modeled here as a function of water content available on the surface.
For evaporation from the surface of the growing medium, this is modelled in this work as
\begin{equation}
\epsilon_\mathrm{evap}({W_\mathrm{med}}) = \begin{cases}
	1 		& \text{if } W_\mathrm{med\%} > k_\mathrm{W_{per}} \\
	k_\mathrm{G_W} W_\mathrm{med\%}& \text{if }  W_\mathrm{med\%} \leq k_\mathrm{W_{per}}, \\
\end{cases}\label{eq:eps_evap}
\end{equation}
where $k_\mathrm{G_W}$ is the conductivity offered by the feed for transport of water to the growing medium surface  and $k_\mathrm{W_{per}}$ is the moisture percentage above which the growing medium is moist enough to allow maximum evaporation.

From above Equation~\cref{eq:gen_evap_cond} replacing the terms $k_\mathrm{A_{surf}}$, $ k_\mathrm{h_{surf}}$, $T_\mathrm{surf}$ and $W_\mathrm{med\%}$ with the terms corresponding to actual surfaces, in this case the growing medium, walls and the heat-exchanger, gives the corresponding water flux terms $\phi_\mathrm{W_{L,chm}}$, $\phi_\mathrm{W_{L,TEC}}$ and $\phi_\mathrm{W_{L,med}}$.

Water vapor transport due to the air exchange through ventilation, leakage, and door opening events similar to heat exchange through ventilation, is given by
\begin{equation}\label{eq:h_exc}
	\begin{aligned}
	\phi_\mathrm{H_{exch}} 	&=\phi_\mathrm{\dot{V}_u} (H_\mathrm{out} - H_\mathrm{air}), \\
	\phi_\mathrm{H_{leak}} 	&= k_\mathrm{\dot{V}_{leak}} (H_\mathrm{out} - H_\mathrm{air}), \\
	\phi_\mathrm{H_{door}} 	&= u_\mathrm{d} k_\mathrm{\dot{V}_{door}} (H_\mathrm{out} - H_\mathrm{air}).
	\end{aligned}
	\end{equation}

Finally, the water influx to the growing medium (in liquid form) is through pumps and depend on the actuator properties given as
\begin{equation}\label{eq:w_u}
	\phi_\mathrm{W_{u}} = k_\mathrm{W_u} u_\mathrm{W_{med}}
\end{equation}
where $k_\mathrm{W_u}$ is the water transport rate of the pumps.


\paragraph{Air composition:}
Metabolic activity of larvae consumes O$_2$ and feed, producing CO$_2$ as a result.
The concentration of the two gases O$_2$ and CO$_2$ change relative to each other and modeling both these concentrations may not be necessary when air is supplied from natural source with fixed O$_2$ to CO$_2$ ratio.
The gas concentration changes and the flux due to ventilation, leakage, and door opening events are given respectively as
\begin{equation}\label{eq:phi_cexch}
	\begin{aligned}
		\phi_\mathrm{C_{exch}} &= \phi_\mathrm{\dot{V}_u}(C_\mathrm{out}-C_\mathrm{air}), \quad \phi_\mathrm{O_{exch}} = \phi_\mathrm{\dot{V}_u}(O_\mathrm{out}-O_\mathrm{air}), \\
		\phi_\mathrm{C_{leak}} &=  k_\mathrm{\dot{V}_{leak}}(C_\mathrm{out}-C_\mathrm{air}), \quad \phi_\mathrm{O_{leak}} =  k_\mathrm{\dot{V}_{leak}}(O_\mathrm{out}-O_\mathrm{air}),\\
		\phi_\mathrm{C_{door}} &=  u_\mathrm{d}k_\mathrm{\dot{V}_{door}}(C_\mathrm{out}-C_\mathrm{air}), \quad \phi_\mathrm{O_{door}} =  u_\mathrm{d}k_\mathrm{\dot{V}_{door}}(O_\mathrm{out}-O_\mathrm{air}).
	\end{aligned}
\end{equation}

\setcounter{figure}{0}
\setcounter{table}{0}

\section{Experiment setup and experiments}\label{app:exp_setup_data}
The production setup used, raw data obtained using the setup, and the additional experiments performed to identify the characteristics of the production setup are described in this section.

\subsection{Production Unit}\label{app:exp_setup}
\begin{figure}[!h]
	\includegraphics[width=\linewidth,trim={150 100 0 180},clip]{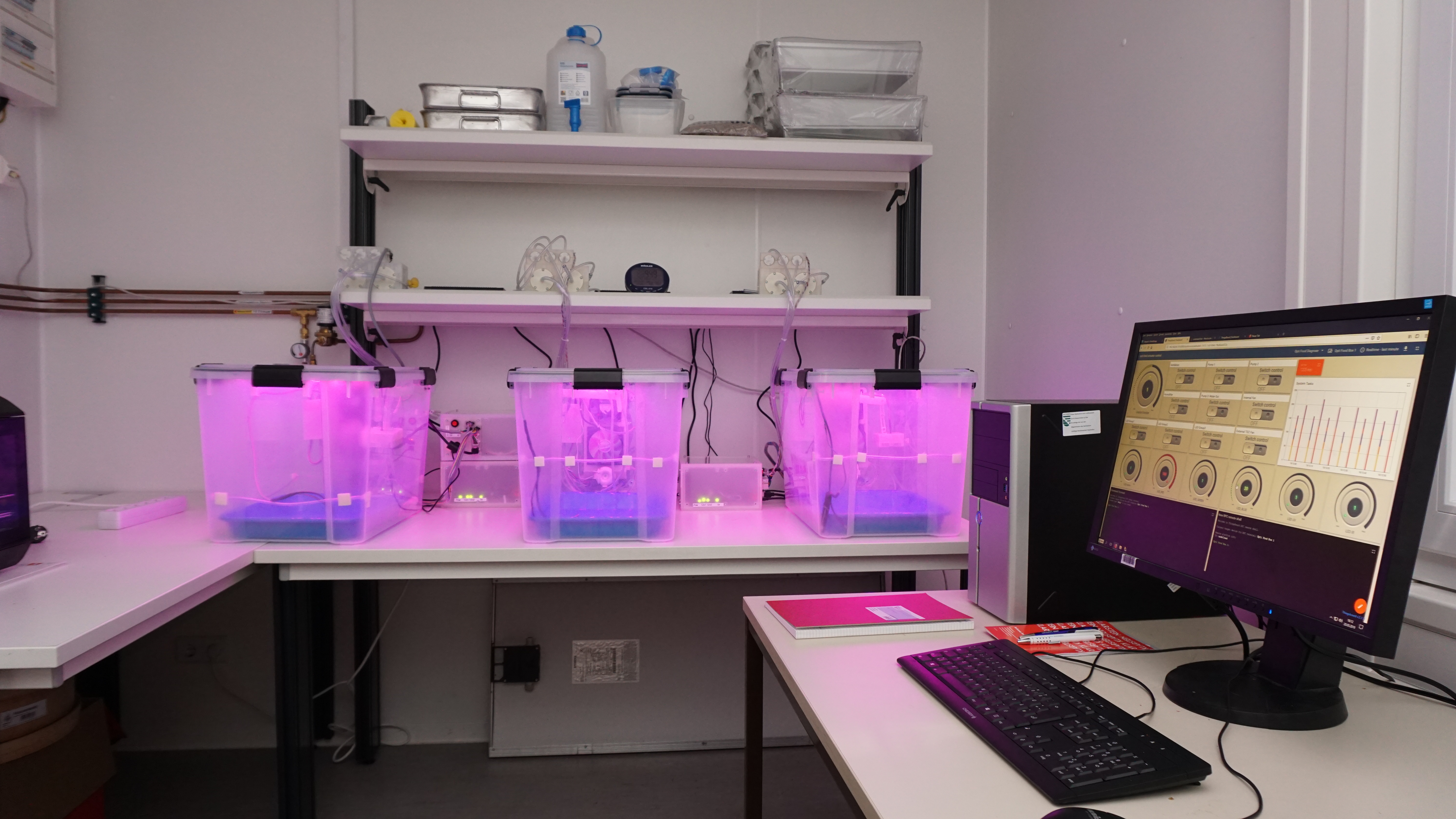}\caption{{\bf Experiment Setup.} Three independent production units, with individual climate control and integrated sensors.}\label{fig:acsd_setup}
\end{figure}

The controlled environment (see Fig~\ref{fig:acsd_setup}) has a volume of \SI{75}{\liter} and holds a growing tray of dimension \SI{22}{\centi\meter} x \SI{32}{\centi\meter} x \SI{5.5}{\centi\meter} and can contain up-to \SI{4}{\kilogram} of substrate.
This growing tray serves as a container for the growing medium that contains selected feed for the larvae, selected number of young larvae (neonates) and the microbiome that eventually develops and grows along with the larvae in the growing medium.
The temperature, humidity, airflow/air-concentration, light intensity and day-night cycles within the unit can be regulated as required.
Information related to the states inside the production unit such as temperature of air and growing medium; CO$_2$ and O$_2$ concentrations; and humidity in air and moisture and temperature in growing medium are recorded by the sensors integrated within.
Similarly, information related to the states outside the production unit, e.g., temperature, humidity and CO$_2$ concentration of external air source, are logged using a data logger. Further details regarding the production environment can be found in our previous work \citep{Padmanabha2019} (see Section 2.1.4 and 3.7).

\subsection{Thermodynamic experiments for characterizing the production setup}\label{app:td_exp}
The production unit is operated in different modes under different conditions as listed in Table~\ref{tab:exp_td}.
In passive mode, the contents (air or growing medium) are preheated and the variations in temperature and humidity are observed.
In active automated mode, the actuators execute pre-programmed sequence to generate specific temperature and ventilation profiles and the resulting temperature and humidity changes are logged.
\begin{table}[h]
		\caption{{\bf Thermodynamics experiment configurations}}
	 	\begin{tabular}{|m{0.1\textwidth}|m{0.25\textwidth}|m{0.25\textwidth}|m{0.15\textwidth}|m{0.1\textwidth}|}
		\hline
		ID & Growing medium & mode & TEC & Ventilator \\\thickhline
		TD1 & none& passive, pre heated air & OFF & OFF\\\hline	
		TD2 &\SI{1}{\kilogram} water covered with conductive film& passive, pre heated growing medium & OFF & OFF\\\hline
		TD3 & none & active, automated & 100\% to -100\%& OFF\\\hline
		TD4 & \SI{1}{\kilogram} water covered with conductive film& active, automated & 100\% to -100\% & OFF\\\hline
		TD5 & \SI{1}{\kilogram} water without conductive film& active, automated & 100\% to -100\% & OFF\\\hline
		TD6 & \SI{1}{\kilogram} water without conductive film& active, automated & 100\% to -100\% & ON\\
		\hline  
	\end{tabular}
	{In the active automated tests, the TEC based heating-cooling system is set to +100\% and allowed to heat for \SI{1}{\hour} to reach temperature saturation.
	The power is then reduced in steps of 5\% and allowed to saturate for \SI{20}{\minute} until the power reaches -100\% and then turned OFF to allow the temperatures to reach the outside temperatures passively.
	These tests cover a wide range of temperatures, transport mechanisms, and flux directions for the heat and water transport.}
 \label{tab:exp_td}
\end{table}

\subsection{Models goodness of fit for heat and water fluxes without larvae}\label{app:td_validation}
The thermodynamic properties of the production unit along with the heat and water fluxes are identified using the data from TD1--TD6 experiments based on the developed models.
This step is important to obtain accurate convective and conductive heat transfer properties of the production unit and necessary to separate the resource and energy flux contributed due to the larvae growth and interaction with its environment.
This identification enables to distinguish between the heat produced by the biological component and the latent heat of evaporation.

\begin{figure}[!h]
	\includegraphics[width=\linewidth,trim={0 0 0 35},clip]{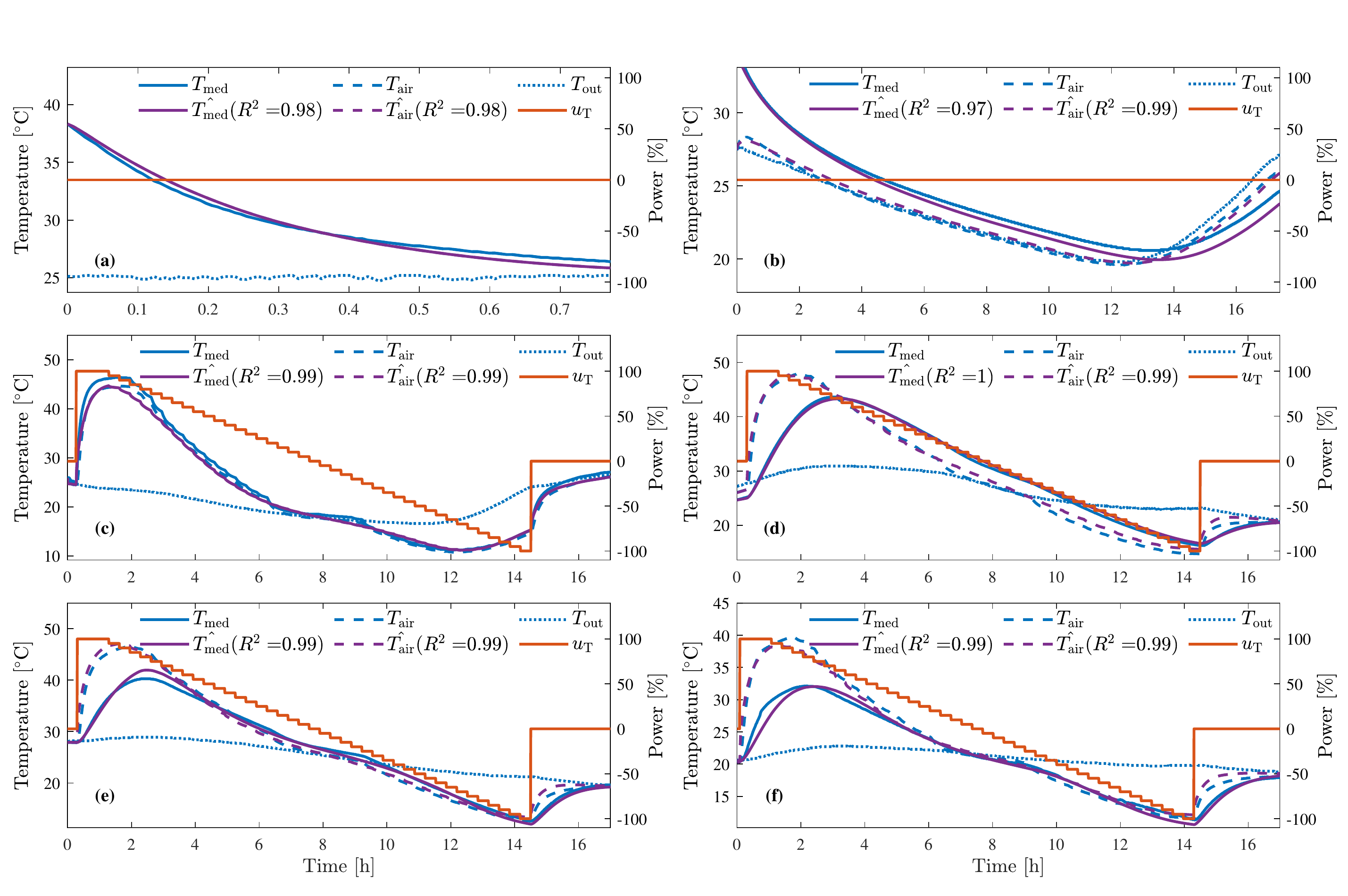}\caption{\textbf{Measurement vs. simulation (entities with hat): dynamic temperature changes under different conditions.}
	\textbf{(a)} TD1: convective losses between $T_\mathrm{air}$ and $T_\mathrm{out}$.
	\textbf{(b)} TD2: convective transfer between $T_\mathrm{air}$ and $T_\mathrm{med}$ only through heat transfer.
	\textbf{(c)} TD3: actuator influenced convective transfer between $T_\mathrm{hx}$, $T_\mathrm{air}$ and $T_\mathrm{out}$.
	\textbf{(d)} TD4: actuator influenced convective transfer between $T_\mathrm{hx}$, $T_\mathrm{air}$ and $T_\mathrm{med}$ only through heat transfer.
	\textbf{(e)} TD5: same as TD4 but through simultaneous heat and mass transfer with heat loss due to latent heat of evaporation.
	\textbf{(f)} TD6: same as TD5 but also additional mass flux between production unit and external environment through ventilation.
	}
	\label{fig:TD_model_valid}
\end{figure}

Models are simulated using the starting values, actuator signals, and disturbance signals as logged in the experiments. The simulation results compared to the actual measurements are illustrated in Fig~\ref{fig:TD_model_valid}.
The convective heat transfer parameters obtained using TD1 and TD2 successfully describe the heat transfer between the production environment, growing medium and external environment (see Fig~\ref{fig:TD_model_valid}\textbf{(a)} and \textbf{(b)}).
Parameters obtained using TD3 describe the dynamics of the heating-cooling system and the heat exchange between the heat exchanger $T_\mathrm{hx}$ and the air $T_\mathrm{air}$.
Comparing the temperature $T_\mathrm{med}$ and $T_\mathrm{air}$ in Fig~\ref{fig:TD_model_valid}\textbf{(d)}, \textbf{(e)}, and \textbf{(f)} in the time period \SI{4}{\hour}--\SI{12}{\hour}, it can be observed that the heat transfer between the ambient air and the growing medium is higher in the experiments TD5 and TD6.
The higher transfer rate is due to increase in convective transfer coefficient due to the vapor transport (evaporation) taking place in TD5 and TD6.
Evaporation of water from the growing medium causes additional loss of heat.
This loss of heat is proportional to the latent heat of evaporation and results in cooling of the growing medium.
This can be inferred from the increasing temperatures difference $T_\mathrm{air}-T_\mathrm{med}$ in TD4--TD6 observed from \SI{0}{\hour}--\SI{3}{\hour} in Fig~\ref{fig:TD_model_valid}\textbf{(d)}, \textbf{(e)}, and \textbf{(f)}.

In conclusion, the derived model of the production environment together with the identified model parameters, accurately describe the functioning of the individual components of the production system with $R^2 > 0.97$.
The only slight deviation observed (see Fig~\ref{fig:TD_model_valid}\textbf{(c)}) is due to the difference in the construction of the sensor housing, where the sensor measuring $T_\mathrm{air}$ has slightly lower sensitivity compared to the sensor measuring $T_\mathrm{med}$.
With these results, it can be stated that the model requirements \textbf{1} and \textbf{3} for the production environment are satisfied.

\subsection{Raw data obtained in larvae growth experiments TG2}\label{app:tg2_raw}
\begin{figure}[!h]
	\includegraphics[width=\linewidth,trim={0 10 0 30},clip]{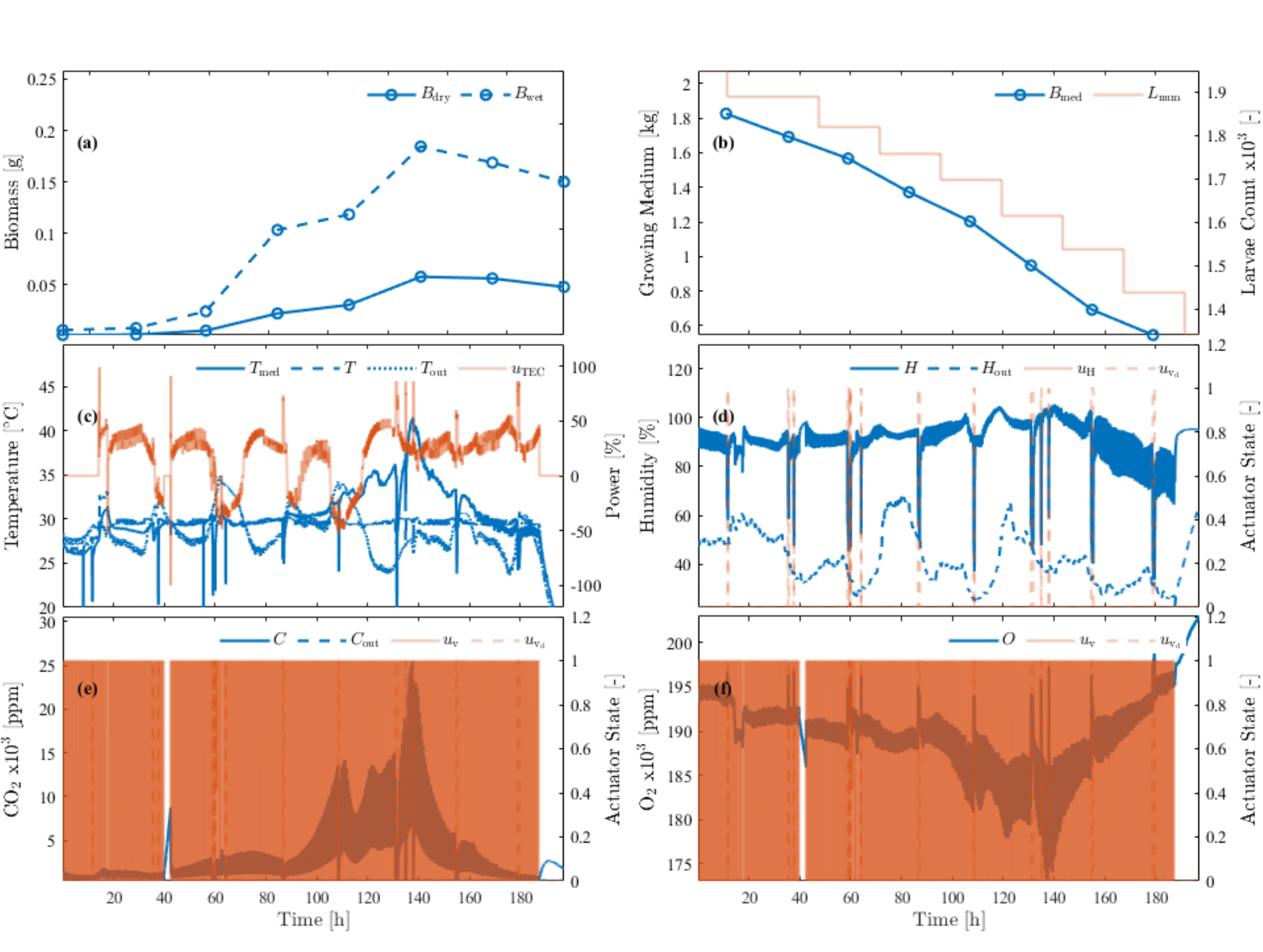}\caption{\textbf{Temperature dependant larvae growth experiment TG2.}Temperature set point of \SI{29}{\celsius} and tracking the development of larvae and resource changes in production unit using sensors and manual measurements.
	 \textbf{(a)} Average wet and dry mass of larva samples.
	 \textbf{(b)} Total mass of the growing medium including larvae, substrate and water.
	 \textbf{(c)},\textbf{(d)},\textbf{(e)} and \textbf{(f)} are temperature, humidity, CO$_2$ concentration, and O$_2$ concentration measurements respectively.}
	\label{fig:T_experiments_data}
\end{figure}

The sensor data appears to be jittery as seen in Fig~\ref{fig:T_experiments_data} and this is simply due to the periodic air exchange represented by $u_\mathrm{v}$.
	 Significant drop in sensor values for a short time period can be observed which is due to the opening of the production unit for extracting samples.
	 Despite the temperature set to \SI{29}{\celsius}, increase in substrate temperature can be seen in \textbf{(c)}. This increase in temperature and also the CO$_2$ production are the byproducts of the metabolic activities of the larvae and the microbiome.

\setcounter{figure}{0}
\setcounter{table}{0}

\section{Model parameters} \label{app:model_params}
The Table~\ref{tab:syms_list} lists all the constants and model parameters identified for both the larvae growth dynamics model and the resource and energy flux models.

\sisetup{round-precision=2,round-mode=figures,scientific-notation=engineering}
	\begin{longtable}{p{0.07\textwidth}p{0.65\textwidth}p{0.1\textwidth}p{0.1\textwidth}}
		\caption{\bf List of all constants and parameters.} \label{tab:syms_list} \\ 
		\hline	\bf	Symbol                     & \bf Description                                		&\bf Value& \bf Unit                  \\ \hline \endfirsthead
		\caption[]{(continued ...)} \\
		\hline	\bf	Symbol                     & \bf Description                                		&\bf Value& \bf Unit                  \\ \hline 	\endhead
		\multicolumn{3}{r}{{continued on next page ...}} \\ \hline	\endfoot
		\hline \hline \endlastfoot
		\multicolumn{3}{l}{Larvae model parameters}\\
		$k_\mathrm{inges}$ 	& specific ingestion rate per larva& \num{37.39e-6} &[\si{\gram\per\gram\per\second}] \\
		$k_\mathrm{mat}$ 	& specific maturity rate per larva& \num{1.6540e-05}&[\si{\gram\per\gram\per\second}] \\
		$k_\mathrm{maint}$ 	& specific maintenance rate per larva& \num{2.1690e-06}&[\si{\gram\per\gram\per\second}] \\
		$k_\mathrm{\alpha_{excr}}$ 	& fraction of ingested feed excreted out & 0.25&[-] \\
		$k_\mathrm{\alpha_{assim}}$ & fraction of ingested feed spent for digestion& 0.1843&[-] \\
		$k_\mathrm{T_{\Sigma}1}$  	& development sum at which the assimilation starts to cease and maturity starts& 261&[\si{\hour}] \\
        $k_\mathrm{T_{\Sigma}2}$  	& development sum at which the assimilation process ends & 272&[\si{\hour}] \\
		$k_\mathrm{T_{\Sigma}3}$  	& development sum at which the maturity process end & 286&[\si{\hour}] \\
		$k_\mathrm{Q_{assim}}$  	& specific heat production through assimilation respiration & \num{0.01401e6}&[\si{\joule\per\gram}] \\
		$k_\mathrm{Q_{mat}}$  	& specific heat production through maturity respiration & \num{0.02802e6}&[\si{\joule\per\gram}] \\
		$k_\mathrm{Q_{miant}}$  	& specific heat production through maintenance respiration & \num{0.02802e6}&[\si{\joule\per\gram}]\\
		$k_\mathrm{Q_{bio}}$  	& specific heat production through microbiome respiration & \num{3.1e3}&[\si{\joule\per\gram}] \\
		$k_\mathrm{W_{assim}}$  	& specific water consumption per gram feed assimilated by the larvae & \num{2.9}&[\si{\gram\per\gram}] \\
		$k_\mathrm{C_{assim}}$  	& specific CO$_2$ production through assimilation respiration & \num{1.6}&[\si{\gram\per\gram}] \\
		$k_\mathrm{C_{mat}}$  	& specific CO$_2$ production through maturity respiration & \num{1.6}&[\si{\gram\per\gram}] \\
		$k_\mathrm{C_{miant}}$  	& specific CO$_2$ production through maintenance respiration & \num{1.6}&[\si{\gram\per\gram}] \\
		$k_\mathrm{C_{bio}}$  	& specific CO$_2$ production through microbiome respiration & \num{1.1920e-04}&[\si{\gram\per\gram}] \\
		$k_\mathrm{bio_{C:O}}$  	& O$_2$ consumed for every CO$_2$ produced & \num{1}&[-] \\ \\
		\multicolumn{4}{l}{Production unit specific model parameters}\\
		$k_\mathrm{V_{chm}}$ 		& total volume inside the production unit 		& \num{0.064}&[\si{\meter\cubed}] \\
		$k_\mathrm{A_{c}}$ 		& total surface area of the production unit & \num{1.117}&[\si{\meter\squared}] \\
		$k_\mathrm{A_{m}}$ 		& total surface area of the growing medium & \num{0.12}&[\si{\meter\squared}] \\
		$k_\mathrm{A_{hx}}$ 	& total surface area of the heat exchanger unit & \num{0.29}&[\si{\meter\squared}] \\
		$k_\mathrm{h_{a-c}}$ 	& convective heat transfer coefficient for air-wall interface & \num{26.97}&[\si{\watt\per\meter\squared\per\kelvin}] \\
		$k_\mathrm{h_{a-m}}$ 	& convective heat transfer coefficient for air-growing medium interface & \num{12.9}&[\si{\watt\per\meter\squared\per\kelvin}] \\
		$k_\mathrm{h_{a-hx}}$ 	& convective heat transfer coefficient for air-heat exchanger interface & \num{26.04}&[\si{\watt\per\meter\squared\per\kelvin}] \\
		$k_\mathrm{h_{a-o}}$ 	& convective heat transfer coefficient for wall-outside air interface & \num{8.46}&[\si{\watt\per\meter\squared\per\kelvin}] \\
		$k_\mathrm{he_{a-m}}$ 	& mass flow absent convective heat transfer coefficient for air-growing medium interface & \num{11.75}&[\si{\watt\per\meter\squared\per\kelvin}] \\
		$k_\mathrm{hm_{a-m}}$ 	& mass flow influenced additional convective heat transfer coefficient for air-growing medium interface & \num{1.34}&[\si{\watt\per\meter\squared\per\kelvin}] \\
		$k_\mathrm{A_{hx-c}}$ 	& total surface area of contact at heat exchanger-wall interface & \num{0}&[\si{\meter\squared}] \\
		$k_\mathrm{A_{m-c}}$ 	& total surface area of contact at growing medium-wall interface & \num{0}&[\si{\meter\squared}] \\
		$k_\mathrm{U_{hx-c}}$ 	& conductive heat transfer coefficient for heat exchanger-wall interface & \num{0}&[\si{\watt\per\meter\squared\per\kelvin}] \\
		$k_\mathrm{U_{m-c}}$ 	& conductive heat transfer coefficient for growing medium-wall interface & \num{0}&[\si{\watt\per\meter\squared\per\kelvin}] \\
		$k_\mathrm{c_{air}}$ 		& specific heat capacity of air             & \num{1006}&[\si{\joule\per\kelvin\per\kilogram}] \\
		$k_\mathrm{\rho_{air}}$ 	& density of air             				& \num{1.2041}&[\si{\kilogram\per\meter\cubed}] \\
		$k_\mathrm{\dot{V}_{u}}$ & mass transfer rate through ventilation pumps&\num{2.5e-4} &[\si{\meter\cubed\per\second}] \\
		$k_\mathrm{\dot{V}_{leak}}$ & mass transfer rate through leakage in the production unit &\num{ 5.7870e-07} &[\si{\meter\cubed\per\second}] \\
		$k_\mathrm{\dot{V}_{door}}$ & mass transfer rate through door in the production unit &\num{1.5000e-04} &[\si{\meter\cubed\per\second}] \\
		$k_\mathrm{a,q}$      	& Seebeck coefficient of the TEC element &\num{0.0460} &[\si{\volt\per\kelvin}] \\
		$k_\mathrm{V_{max}}$ 		& maximum applicable voltage to TEC & \num{12}&[\si{\volt}] \\
		$k_\mathrm{R_{q}}$ 		& Internal resistance of the TEC  element& \num{1.72}&[\si{\ohm}] \\
		$k_\mathrm{TEC}$ 		& Thermal conductivity of the TEC  element& \num{0.4224}&[\si{\watt\per\kelvin}] \\
		$k_\mathrm{c_{water}}$		& specific heat capacity of water     		&\num{4182} &[\si{\joule\per\kelvin\per\kilogram}] \\
		$k_\mathrm{h_{ew}}$			& latent heat of vaporization of water			& \num{2256400}&[\si{\kilo\joule\per\kilogram}] \\		
		$k_\mathrm{c_{tray}}$		& specific heat capacity of growing tray     		& \num{500}&[\si{\joule\per\kelvin\per\kilogram}] \\
		$k_\mathrm{m_{tray}}$		& mass of growing tray     		& \num{0.85}&[\si{\kilogram}] \\
		$k_\mathrm{c_{feed}}$		& specific heat capacity of feed (dry mass)     & \num{19861}&[\si{\joule\per\kelvin\per\kilogram}] \\
		$k_\mathrm{c_{vap}}$		& specific heat capacity of water vapor    & \num{4182}&[\si{\joule\per\kelvin\per\kilogram}] \\	
		$k_\mathrm{c_{chm}}$		& specific heat capacity of chamber internal parts  & \num{1854}&[\si{\joule\per\kelvin\per\kilogram}] \\
		$k_\mathrm{W_{mmol}}$	& molar mass of water & \num{611}&[\si{\gram\per\mole}] \\ 
		$k_\mathrm{R_g}$	& gas constant &\num{461.5} &[\si{\joule\per\kelvin\per\mole}] \\
		$k_\mathrm{G_W}$ &conductivity of the feed for transport of water to the surface &\num{1}&[-]\\
		$k_\mathrm{W_{per}}$ &moisture percentage above which maximum evaporation is possible&0.6&[-]\\
		$k_\mathrm{W_u}$ & water transport rate of the pumps &\num{0.0038}& [\si{\kilogram\per\second}]\\
	\end{longtable}


\section*{Acknowledgments}
This measure is co-financed with tax revenues on the basis of the budget passed by the Saxon state parliament (grant number SAB 100403339; project NutriCon) and of the Federal Ministry of Education and Research of Germany (BMBF; grant number 031B0733D; project CUBEScircles).

\bibliography{bib/references}	

\end{document}